\documentclass{emulateapj}
\begin{document}

\title{High Resolution Imaging of Warm and Dense Molecular Gas 
in the Nuclear Region of the Luminous Infrared Galaxy NGC~6240}
\author {
Daisuke Iono\altaffilmark{1},
Christine D. Wilson\altaffilmark{2,3},
Shigehisa Takakuwa\altaffilmark{1},
Min S. Yun\altaffilmark{4},
Glen R. Petitpas\altaffilmark{2},
Alison B. Peck\altaffilmark{2},
Paul T. P. Ho\altaffilmark{2,6},
Satoki Matsushita\altaffilmark{6},
Ylva M. Pihlstrom\altaffilmark{5},
Zhong Wang\altaffilmark{2}
}
\altaffiltext{1}{National Astronomical Observatory of Japan, 2-21-1 Osawa, Mitaka, Tokyo 181-0015, Japan; d.iono@nao.ac.jp}
\altaffiltext{2}{Harvard-Smithsonian Center for Astrophysics, 60 Garden Street, Cambridge, MA 02138}
\altaffiltext{3}{Department of Physics and Astronomy, McMaster University, Hamilton, ON L8S 4M1, Canada}
\altaffiltext{4}{Department of Astronomy, University of Massachusetts, Amherst, MA 01003}
\altaffiltext{5}{Department of Physics and Astronomy, University of New Mexico, 800 Yale Boulevard NE, Albuquerque, NM 87131}
\altaffiltext{6}{Academia Sinica Institute of Astronomy and Astrophysics, P.O. Box 23-141, Taipei 106, Taiwan, R.O.C.}

\begin{abstract}
We present $\sim 2''$ resolution 
CO~(3--2), HCO$^+$(4--3) and $880\micron$ continuum images of the 
luminous infrared galaxy NGC~6240 obtained at the Submillimeter Array.
We find that the 
spatially resolved CO~(3--2), HCO$^+$(4--3) and the $880\micron$ 
emission peaks between the two nuclear components that are 
both known to harbor AGNs. 
Our Large Velocity Gradient (LVG) analysis performed on each 
velocity channel suggests that the peak of the molecular gas 
emission traced in our observations is warm (T = 20 -- 100~K), 
dense ($n_{\rm H_2} = 10^{5.0 - 5.4}$~cm$^{-3}$) and moderately optically thin 
($\tau = 0.2$ -- 2) in the central 1~kpc. We also 
find large column densities of $\sim 10^{23}$~cm$^{-2}$.
Such extreme conditions are observed over $\sim 300$~km~s$^{-1}$ centered
around the CO derived systemic velocity.
The derived molecular gas mass from the CO~(3--2) emission and a 
CO-to-H$_2$ conversion factor commonly used for ULIRGs is 
$(6.9 \pm1.7) \times 10^{9}$~M$_{\odot}$, and this is consistent with the
mass derived from previous CO~(2--1) observations. 
The gas is highly turbulent in the central kpc
($\Delta v_{\rm FWZI} \sim 1175$~km~s$^{-1}$).
Furthermore, possible inflow or outflow activity is suggested 
from the CO~(3--2) velocity distribution. 
We tentatively state that $3.5 \times 10^8$~M$_{\odot}$ of 
isolated CO~(3--2) emission seen west of the northern disk 
may be associated with outflows from starburst superwinds, 
but the gas outflow scenario from one of the central AGN is not 
completely ruled out.  Piecing all of the information together, 
the central region of NGC~6240 harbors 2 AGNs, 
$\sim 10^{10}$~M$_{\odot}$ of molecular gas mass, 
$5 \times 10^7$~M$_{\odot}$ of dust mass,
 and has possible evidence of inflow and outflow activity.
\end{abstract}


\section{Introduction}

(Ultra) Luminous Infrared Galaxies [(U)LIRGs] are systems discovered in the 
local universe that emit extremely large luminosity in the far infrared 
\citep[see][for review]{sanders96,lonsdale06}.
Their disturbed optical/NIR morphologies 
\citep[e.g.][]{armus87,surace98,farrah01,bushouse02} and the highly compact  
molecular gas emission \citep{downes98} suggest that collisions/mergers
of gas rich galaxies and subsequent radial inflow of gas toward the
central region is primarily responsible for triggering extensive bursts
of nuclear star formation \citep{scoville00}, 
feeding the obscured central AGN \citep{imanishi06}, 
or most likely, both \citep{genzel98}. 
Such inflow activity was already predicted in early numerical simulation 
studies that involve collisions of two massive 
gas rich galaxies \citep{barnes96,mihos96,iono04a}.  
More recent simulation studies that include a more realistic set of
gas physics and the evolution of a central black hole 
have tied the merger event with radial gas inflow, central starbursts, 
black hole feeding, and a powerful gas 
outflow \citep{springel05,hopkins05,dimatteo05,narayanan06}. 
These simulations have predicted a self-consistent evolutionary 
scenario from mergers to ULIRGs with a brief optical QSO 
phase \citep{hopkins06}. 

High resolution observations of molecular gas tracers in (U)LIRGs and 
have been limited to low $J$ transition CO emission obtained at mm 
interferometers.  Past single dish measurements of dense gas tracers
such as HCN \citep{gao04} have found that molecular gas in (U)LIRGs is 
highly compressed and dense.  Thus, obtaining sensitive and high 
spatial resolution images of high excitation molecular gas emission  
will provide extremely valuable information regarding the 
star formation process during a merger
and will further constrain the physics implemented in the evolution 
simulations.  In addition, recent surveys of molecular gas traced in
high $J$ transition CO in high-$z$ galaxies \citep{greve05,tacconi06} will 
allow us to conduct a direct comparison of the gas properties 
with local counterparts.

Here we present interferometric CO~(3--2), HCO$^+$(4--3), and $880\micron$
continuum observations of 
the luminous infrared galaxy (LIRG) NGC~6240 obtained at the SMA \citep{ho04}.
Both molecular lines are known to trace 
warmer and denser molecular gas than the commonly observed mm lines such 
as CO~(1--0) or CO~(2--1).
Past SMA observations toward infrared bright galaxies  
suggest that the distribution of the 
CO~(3--2) emission is significantly
different from the lower transition lines of CO \citep{iono04b, wang04}. 

\subsection{Multi-Wavelength Properties of NGC~6240}

NGC~6240 is an infrared luminous merger 
(L$_{\rm FIR} = 3.5 \times 10^{11}$~L$_{\odot}$) \citep{yun02}
at a distance of D$_L$ = 107~Mpc 
($1'' = 493$~pc)\footnote{Adopting H$_0$ = 70~km~s$^{-1}$~Mpc$^{-1}$, 
$\Omega_M$ = 0.3, $\Omega_{\Lambda}$ = 0.7}.  Its disturbed optical morphology
and the compact double nuclei suggest that NGC~6240 is a late stage merger. 
The presence of AGN and extended starbursts were already 
suggested from early X-ray observations using a variety of spaceborne 
instruments \citep[i.e.][and references therein]{ikebe00}.
More recently, \textit{Chandra} has resolved the central region into
two AGNs separated by $\sim 1''$ \citep{komossa03}, 
providing significant evidence that NGC~6240 is a merger
of two AGN host galaxies.  Recent \textit{XMM-Newton} spectroscopy has 
revealed a significant contribution of starburst activity to 
the extended X-ray emission \citep{boller03,netzer05}.

Optical and infrared observations using ground based
and spaceborne telescopes have also been used to study this galaxy intensively.
\textit{HST} and \textit{Keck} imaging have resolved each nucleus into
several distinct components \citep{gerssen04,max05} with an apparently
dusty central region.  A large velocity gradient in ionized gas has 
been identified using optical spectroscopy \citep{gerssen04}.
Near IR imaging has also shown the presence of two $K$-band 
nuclei \citep{scoville00,bogdanovic03}.  
Signatures of starbursts appear to dominate the MIR spectrum \citep{lutz03}.
Mid IR spectroscopy using IRS on the \textit{Spitzer Space Telescope}
suggest a minor ($\sim 3$ -- 20\%) contribution of the AGN 
to the bolometric luminosity \citep{armus06}, but mid infrared 
spectra obtained at \textit{Keck} suggest that the dominant power-source
could be explained by starbursts or an AGN \citep{egami06}. 
Optical spectroscopy using \textit{Subaru} 
has revealed significant  H$_2$ emission  
$0\farcs2$ north of the southern nucleus, suggesting that 
the medium between the two nuclei is dominated by starburst super-winds from 
the southern nucleus \citep{ohyama00}.
Finally, a star formation rate of  $61\pm30$~M$_{\odot}$~yr$^{-1}$ is
obtained from a fit to the FIR SED \citep{yun02} and using the
Schmidt law derived in \citet{kennicutt98} 
to covert to the Star Formation Rate (SFR).

Interferometric millimeter studies have identified significant amounts of
molecular gas in NGC~6240 \citep{bryant99,tacconi99,nakanishi05}.  
The bulk of cold and dense gas 
peaks between the two optical/X-ray nuclei, close to the region where
significant H$_2$ emission \citep{ohyama00} is found (Figure~1).
Millimeter continuum emission also peaks in the same region.
Molecular gas kinematics traced in CO~(2--1) emission show
a large velocity gradient between the two nuclei, suggesting that 
$\sim 10^9$~M$_{\odot}$ of molecular gas is in the process of 
accumulating in the new dynamical center between the two nuclei.

\begin{figure}
  \plotone{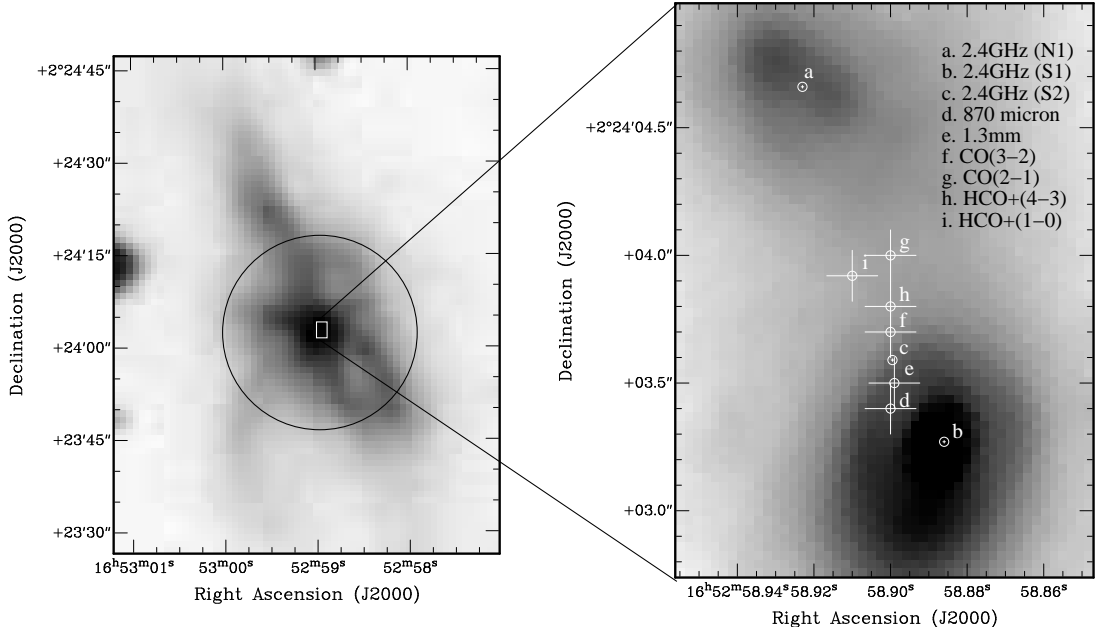}
  \caption{($left$) Digital Sky Survey image of NGC~6240 overlaid 
    with the $35''$ SMA field of view.  ($right$) The \textit{HST} NICMOS 
    $2.2\micron$ image of the central $2''$~(1~kpc) of NGC~6240 in grayscale
    overlaid with various peak positions from radio/mm/submm observations.  
    Since accurate astrometry is not provided in the
    \textit{HST} image, we align the peak of the southern $2.2\micron$
    nucleus with the 2.4~GHz nucleus.
    The length of the crosses indicate 
    the astrometric uncertainties of each measurement.  The points 
    labeled a and b are the locations of the two AGNs.  Coordinates are 
    obtained from (a,b,c) \citet{gallimore04}, (d,f,h) this work, 
    (e,g) \citet{tacconi99}, and (i) \citet{nakanishi05}.
  }
  \label{fig1}
\end{figure}

MERLIN observations \citep{beswick01} have detected 
both AGNs with highly disturbed \ion{H}{1} gas in absorption.  Further,
high resolution VLBA observations \citep{gallimore04} have detected 
two compact radio sources (N1 and S1 in Figure~\ref{fig1}) and an additional 
component (S2) $0\farcs2$ north of the southern nucleus (S1).  
\citet{gallimore04} attribute N1 and S1 to compact radio emission associated 
with the AGNs, and S2 as a radio supernova or material ejected from S1.
An H$_2$O maser detected near S1 \citep{hagiwara03} is attributable to 
either a jet associated with the southern AGN, or with shocks caused by the 
immense collision.  From their low resolution 
VLA maps, \citet{colbert94} have found significant centimeter emission
west of N1.  This emission is resolved into four distinct components, and
they suggest shell material from galactic super-wind as an origin of the
emission.

This article is organized as follows.
In \S2, we present our observational method and the detailed calibration
scheme adopted for the SMA observations.  The CO~(3--2) and HCO$^+$(4--3)
moment maps are presented in \S3, along with the $880\micron$ continuum
image which was made by averaging the emission in the line free channels.  
\S4 focuses on the physical properties of gas.  
We provide a discussion of our results in \S5, and conclude this
paper in \S6.

\section{Observations and Data Reduction}

NGC~6240 was observed on October 9 and 14, 2005 using both the compact 
(unprojected baselines = 16 -- 69 meters) and
extended configurations (unprojected baselines = 50 -- 182 meters) 
of the SMA.  These data were taken in excellent 
atmospheric conditions ($\tau_{230}$ = 0.05 -- 0.06).  
The digital correlator was configured with 2048 channels for the entire 
2~GHz bandwidth for each sideband. 
The receivers were tuned to the redshifted 
CO~(3--2)~($\nu_{rest} = 345.796$~GHz) emission line 
in the lower sideband (LSB), which allows us to simultaneously obtain the
HCO$^+$(4--3)~($\nu_{rest} = 356.734$~GHz) 
line in the upper sideband (USB). The adopted tuning frequency 
was 337.533~GHz~(LSB) using $v_{sys}$ = 7339~km~s$^{-1}$. 
We used $\alpha(\rm J2000) = 16^{\rm h} 52^{\rm m} 58.9^{\rm s}$ and 
$\delta(\rm J2000)= 2^{\circ} 24' 04.00''$ \citep{laurent97}
as our phase reference center.  Initial data calibration 
were carried out in the IDL based SMA calibration package MIR, 
where 3C279, 3C454.3 and Uranus were used for
bandpass calibration.   Absolute flux calibration was performed using
Uranus, and time dependent gain calibration was performed using 
the 1.6~Jy QSO 1743-038 (14~degrees away from NGC~6240) and the 
1.4~Jy QSO 1549+026 (16~degrees away from NGC~6240) as references.

\begin{figure}
  \plottwo{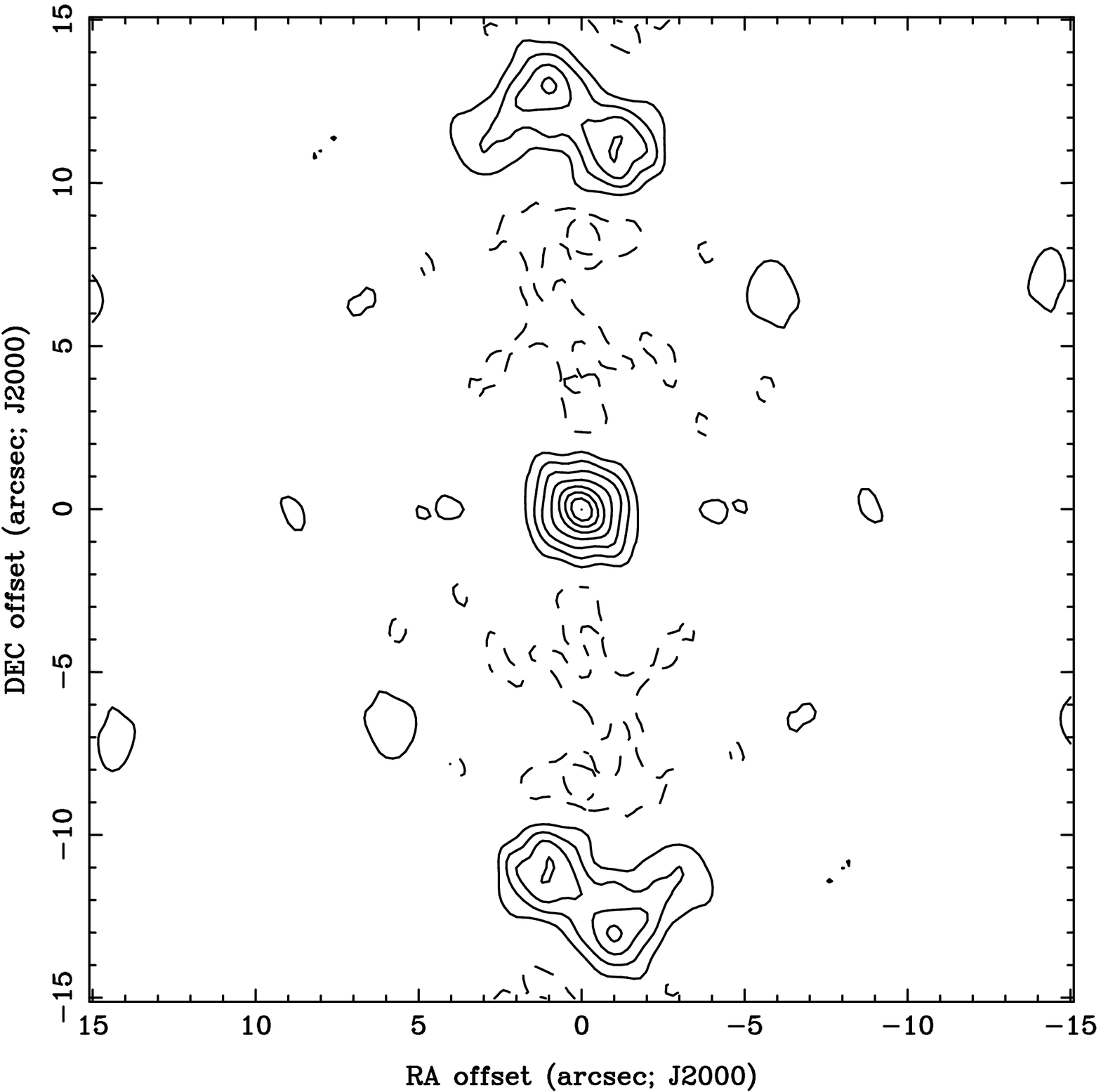}{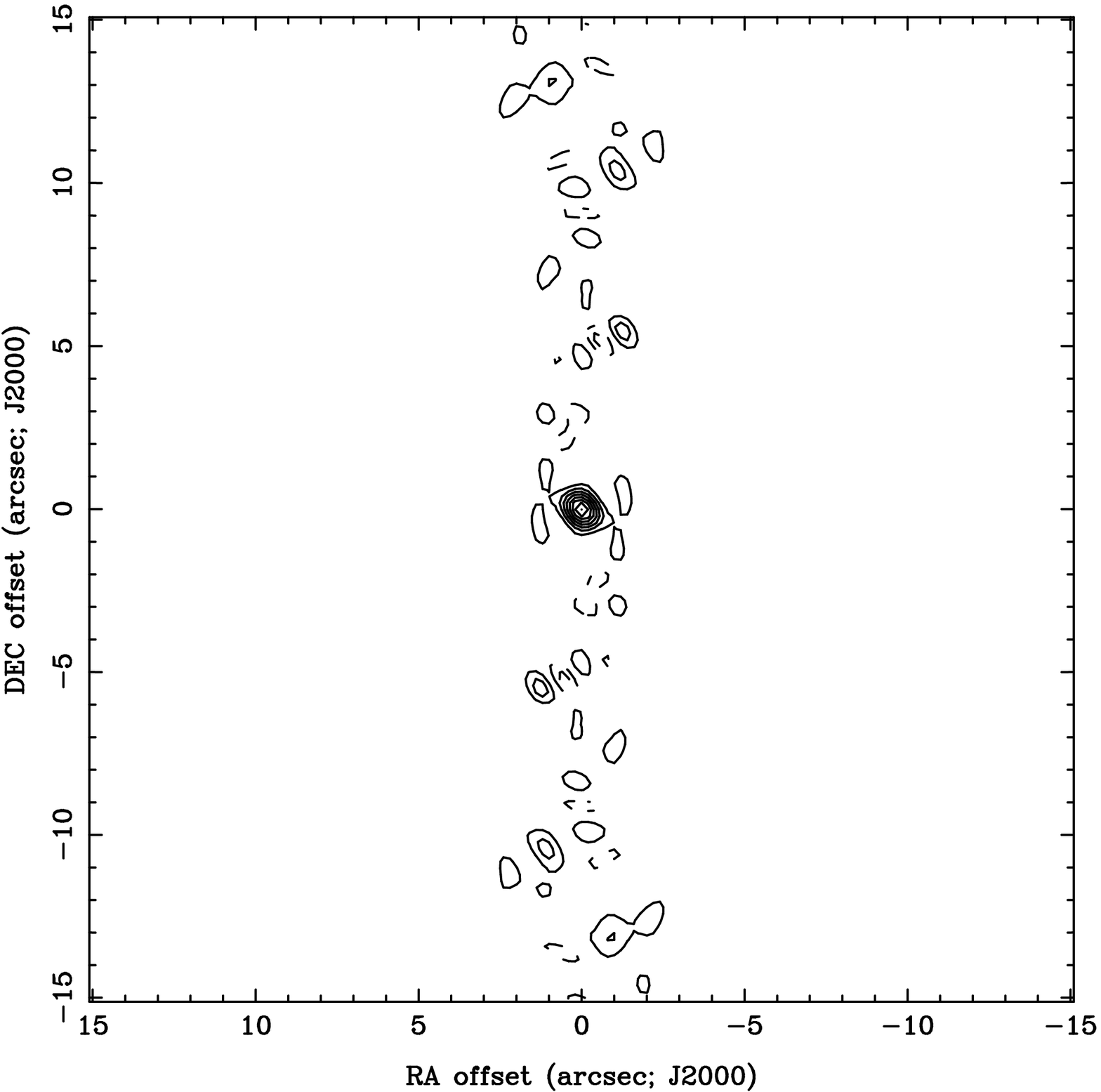}
  \caption{
    The synthesized beam using (\textit{left}) natural and (\textit{right})
    uniform weighting.  The contour levels are 10 -- 90\% of the peak.  
  }
  \label{fig2}
\end{figure}

Synthesis imaging was performed in MIRIAD. 
Channel maps with 25~km~s$^{-1}$ velocity resolution were made 
for the CO~(3--2) line.  We further
varied the visibility weighting from natural (i.e. maximum sensitivity) 
which gave a synthesized beam size of 
$2\farcs0 \times 1\farcs8$~(position angle = 49 degrees; 
Figure~\ref{fig2} (left))  
to uniform (i.e. maximum angular resolution) 
which gave a synthesized beam size of 
$1\farcs1 \times 0\farcs8$~(position angle = 49 degrees; 
Figure~\ref{fig2} (right)).  
Unity weights are assigned for each visibility point in natural weighting, 
giving the best S/N and the coarsest angular resolution.
On the other hand, uniform weighting gives the highest angular resolution
at the cost of lower S/N.  This is achieved by weighting each visibility 
point by the reciprocal of the number of data points found in a grid centered 
around each visibility point.  The rms noise level was 22~mJy and 78~mJy for 
the natural and uniformly weighted maps respectively.  
In addition, in order to derive the peak coordinates of the CO~(3--2) emission,
we have constructed an image using only the longest baseline data which gave 
a synthesized beam size of 
$0\farcs9 \times 0\farcs5$~(position angle = 41~degrees).
For the natural weighted beam, the first sidelobes are significant at 
the 50\% level $\sim 11''$ north and south of the main synthesized beam.
Sidelobes at 10\% level are seen in various positions, the effect of 
which should be negligible with proper deconvolution.
For the uniformly weighted beam, the first significant ($\sim 30\%$) 
sidelobes are located $\sim 5''$ north-south of the main beam. 
CLEANing was performed down to 1.5 times the rms noise level of 
each channel map. For the deconvolution algorithm, we chose CLEAN 
instead of the Maximum Entropy Method (MEM) since the emission is strong and
relatively compact with both the natural and uniformly weighted beams.

The HCO$^+$(4--3) channel maps were made using natural weighting, which 
gave a $2\farcs0 \times 1\farcs7$~(position angle = 42 degrees) 
beam. We note that the HCO$^+$(4--3) emission occurs near the edge of the 
bandpass, and these observations may be missing flux at the low velocity end.
Continuum subtraction was performed by removing  a linear baseline in 
the visibilities in both sidebands.
The 880~$\micron$ 
continuum image was made by adding all of the line free channels in 
the bandpass of both sidebands.  The rms 
noise level for the continuum map was 6~mJy.  From the S/N of the maps,
we estimate the astrometric accuracy of the CO~(3--2) maps presented here 
to be $< 0\farcs1$, and $\sim 0\farcs1$ for the HCO$^+$(4--3) 
and 880~$\micron$ maps.

\section{Results}

The channel maps of CO~(3--2) emission are provided in Figure~\ref{fig3}.
Figure~\ref{fig4}~(left) shows the integrated intensity map of the 
natural weighted CO~(3--2) emission, and Figure~\ref{fig4}~(right) shows
the uniformly weighted image of the central $6''$.  
Figure~\ref{fig5} show the velocity distribution with the same angular 
resolution as in Figure~\ref{fig4}.  The velocity distribution of 
the western complex (WC hereafter; see below) alone is shown in 
Figure~\ref{fig6}.  The integrated intensity maps are clipped at 1.5 times 
the rms in each channel, whereas the velocity distribution maps are clipped 
at 3 times the rms.  The position velocity maps obtained 
along the two AGNs are shown in Figure~\ref{fig7}.
Figure~\ref{fig8} shows the HCO$^+$(4--3) and $880\micron$  
emission images of NGC~6240.  A summary of the derived parameters
is shown in Table~1.

\subsection{CO~(3--2) Emission}

\subsubsection{Distribution}

The size of the CO~(3--2) emission region in the integrated 
intensity map of Figure~\ref{fig4}~($left$) extends 4.4~kpc in both 
north-south and east-west directions.
The coordinates of the peak determined from the highest resolution map
obtained by just using data in the extended configuration are
$\alpha(\rm J2000) = 16^{\rm h} 52^{\rm m} 58.9^{\rm s}$ and 
$\delta(\rm J2000)= 2^{\circ} 24' 03\farcs70$.
This is coincident with the radio component (S2) at 
$\alpha(\rm J2000) = 16^{\rm h} 52^{\rm m} 58.8994^{\rm s}$  and
$\delta(\rm J2000) = 2^{\circ} 24' 03\farcs5925$ 
(error~$= 0.004''$)~\citep{gallimore04} to within $0\farcs1$ 
which is comparable to the astrometric accuracy of the SMA.  
While the inner region of the map shown in Figure~\ref{fig4}~(left) is 
not spatially resolved with this map,
there are several distinct $\sim 1$~kpc scale features that dominate
the emission in the periphery; (1) the emission that extends to the east, 
(2) the emission that extends to the north, and (3) the 
isolated emission west of the main CO~(3--2) concentration (WC).

The $\sim 2$~kpc eastern extension is significant in the velocity range 
of 6964 to 7489~km~s$^{-1}$ (see Figure~\ref{fig3}), and appears to originate
from the southeastern edge of the central emission region.  On the other hand,
the $\sim 2$~kpc northern extension is significant in the 
velocity range 7264 to 7564~km~s$^{-1}$, and appears to display a slight
curve toward the east.  The isolated component (WC) 
was detected $\sim 9''$~(4.4~kpc) west of the 
main CO~(3--2) peak, and this emission is significant in the 
velocity range 7114 to 7714~km~s$^{-1}$.  No significant optical 
counterpart is associated with WC.  
In addition, none of the past interferometric CO observations have 
identified a similar feature \citep{bryant99,tacconi99,nakanishi05}, but
radio continuum observations have detected significant extension of 
synchrotron emission in the same region \citep{colbert94}.

The total integrated intensity from the full resolution map 
including WC is $2834 \pm 20$~Jy~km~s$^{-1}$, recovering
90\% of the total flux measured using a single dish telescope 
\citep[$3180\pm460$~Jy~km~s$^{-1}$;][]{greve06}.
Using the CO--H$_2$ conversion factor appropriate for ULIRGs 
\citep[i.e. $X$~=~$0.8\pm0.2$~M$_\odot$~(K~km~s$^{-1}$)$^{-1}$~pc$^2$;][]{downes98},
and assuming L$^{'}_{\rm CO(3-2)}$~=~L$^{'}_{\rm CO(1-0)}$, 
L$^{'}_{\rm CO(3-2)} = (8.6 \pm 0.1) \times 10^9$~K~km~s$^{-1}$~pc$^2$   
translates to M$_{\rm H_2} = (6.9 \pm 1.7) \times 10^9$~M$_\odot$.
The peak of the CO~(3--2) emission gives a peak H$_2$ column density
of N$_{\rm H_2} = (2.0 \pm 0.3) \times 10^{23}$~cm$^{-2}$ using 
$X$~=~0.8~M$_\odot$~(K~km~s$^{-1}$)$^{-1}$~pc$^2$. 
The peak brightness temperature observed with the
$\sim 2''$ beam is 8.6~K.
The integrated intensity of WC alone is $(143 \pm 20)$~Jy~km~s$^{-1}$, 
which translates to M$_{\rm H_2} = (3.5 \pm 1.0) \times 10^{8}$~M$_\odot$.

The higher resolution image afforded by adopting uniform weighting  
shown in Figure~\ref{fig4}~($right$) resolves out the extended structure,
revealing the compact CO~(3--2) emission that dominates the central 1.5~kpc.
The integrated intensity derived from this map is $1839\pm20$~Jy~km~s$^{-1}$ 
and this is 65\% of the total intensity derived from the full resolution image.
The integrated intensity translates to 
M$_{\rm H_2} = (4.5 \pm 1.1) \times 10^9$~M$_\odot$ assuming 
$X$~=~$0.8\pm0.2$~M$_\odot$~(K~km~s$^{-1}$)$^{-1}$~pc$^2$.
The CO~(3--2) emission arises primarily in a compact structure
centered between the two nuclei and slightly extended in the 
direction defined by the line connecting the two.

\subsubsection{Kinematics}

\begin{figure}
  \plotone{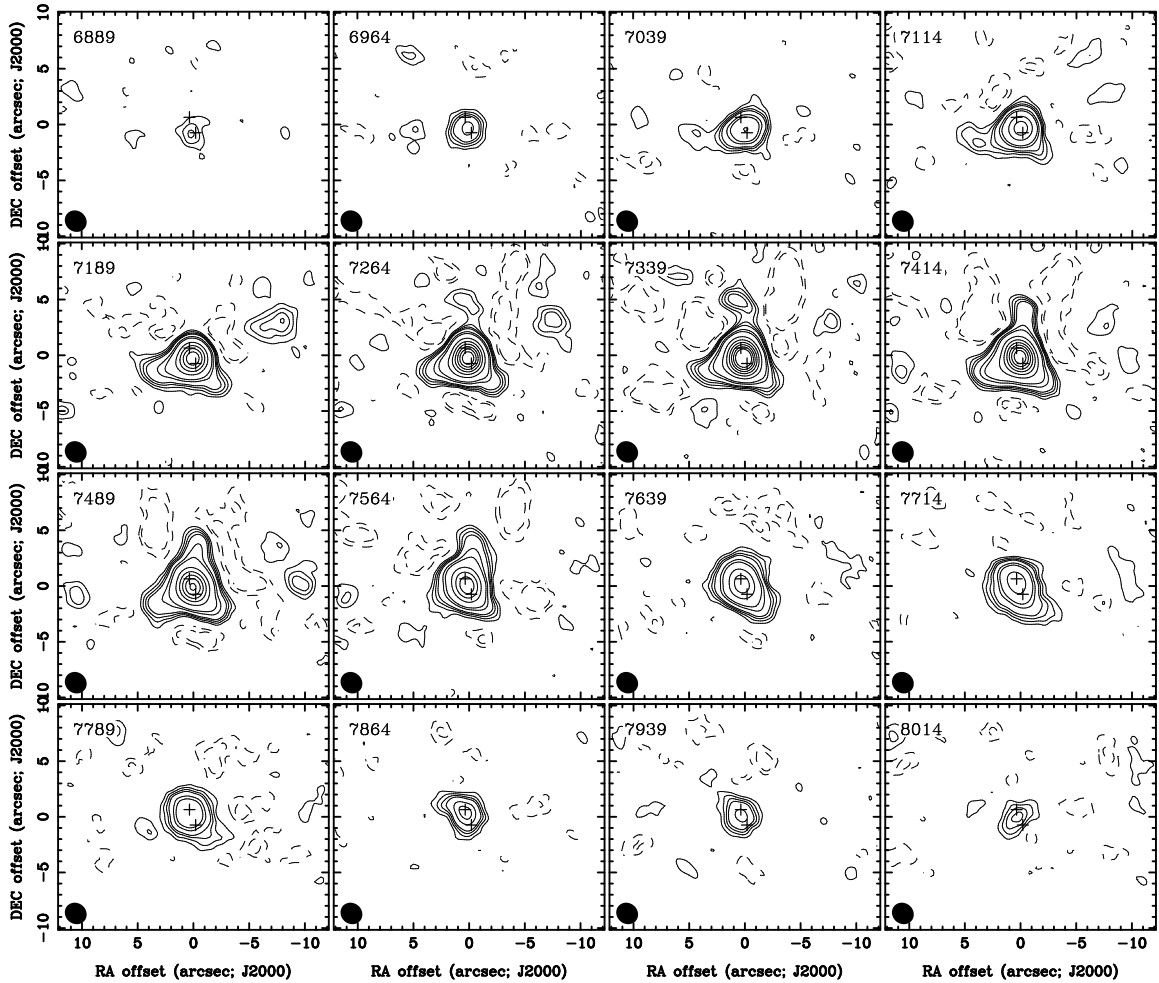}
  \caption{
    The natural weighted CO~(3--2) channel maps in the velocity range 
    6889 to 8014~km~s$^{-1}$.  Although the original channel maps were
    made with 25~km~s$^{-1}$ velocity resolution, the maps presented here
    are made by averaging three channels (75~km~s$^{-1}$) in order to 
    present the extremely broad velocity width in a single figure. 
    The velocities are shown in the upper 
    left corner of each panel, and the synthesized beam is shown in the 
    lower left.  The two crosses show the location of the 
    two AGNs (N1 and S1 in Figure~\ref{fig1}).  The contour levels are 
    16~mJy~(1~sigma)~($\times$ -5,-3,3,5,7,9,15,20,40,60,80,100,120,140).
}
  \label{fig3}
\end{figure}

\begin{figure}
  \plottwo{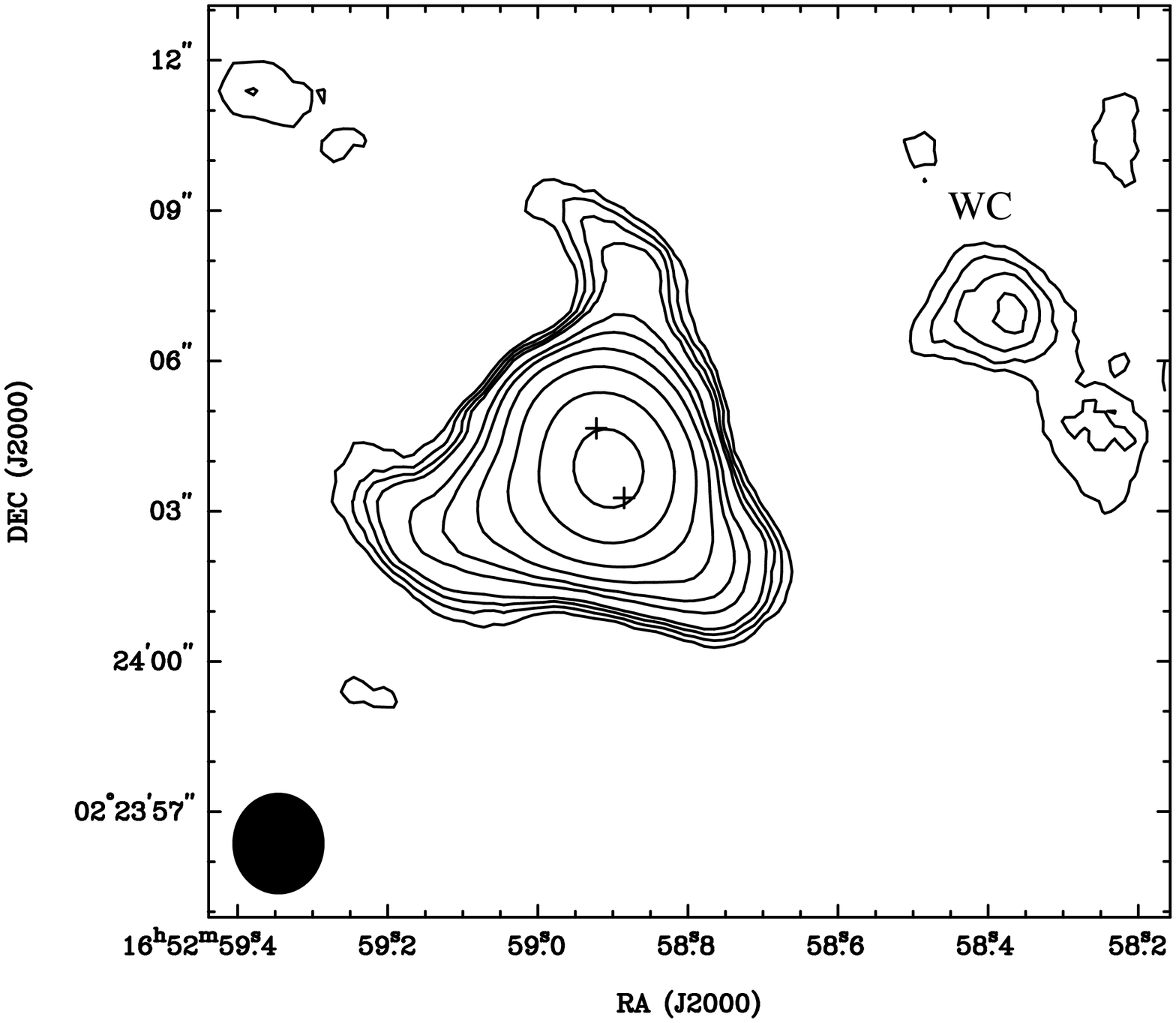}{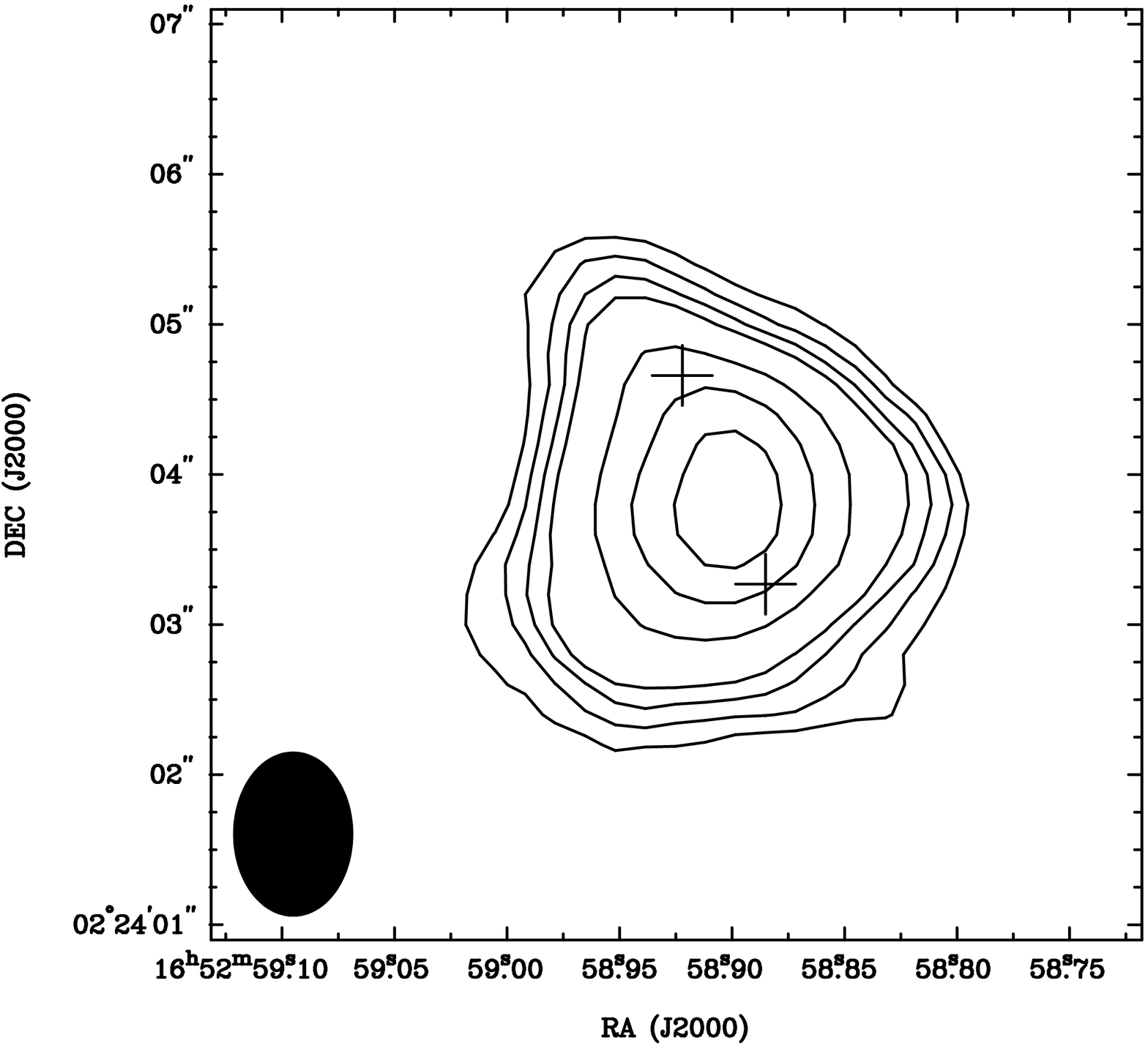}
  \caption{
    ($left$) The CO~(3--2) moment 0 map of the central $18''$ of NGC~6240.  
    The contour levels are 
    5.5~Jy~km~s$^{-1}$~($\times$ 3,5,7,9,15,20,30,50,100,200).  We define the 
    isolated CO~(3--2) emission complex west of the main CO~(3--2) 
    concentration as WC.  ($right$) The uniformly weighted CO~(3--2) moment 
    0 map of the central $6''$ of NGC~6240.  The contour levels are 
    16~Jy~km~s$^{-1}$~($\times$ 3,5,7,9,15,21,27).
    The crosses are the locations of N1 and S1 (see Figure~\ref{fig1}), and
    the synthesized beams are shown in the lower left.
}
  \label{fig4}
\end{figure}

\begin{figure}
  \plottwo{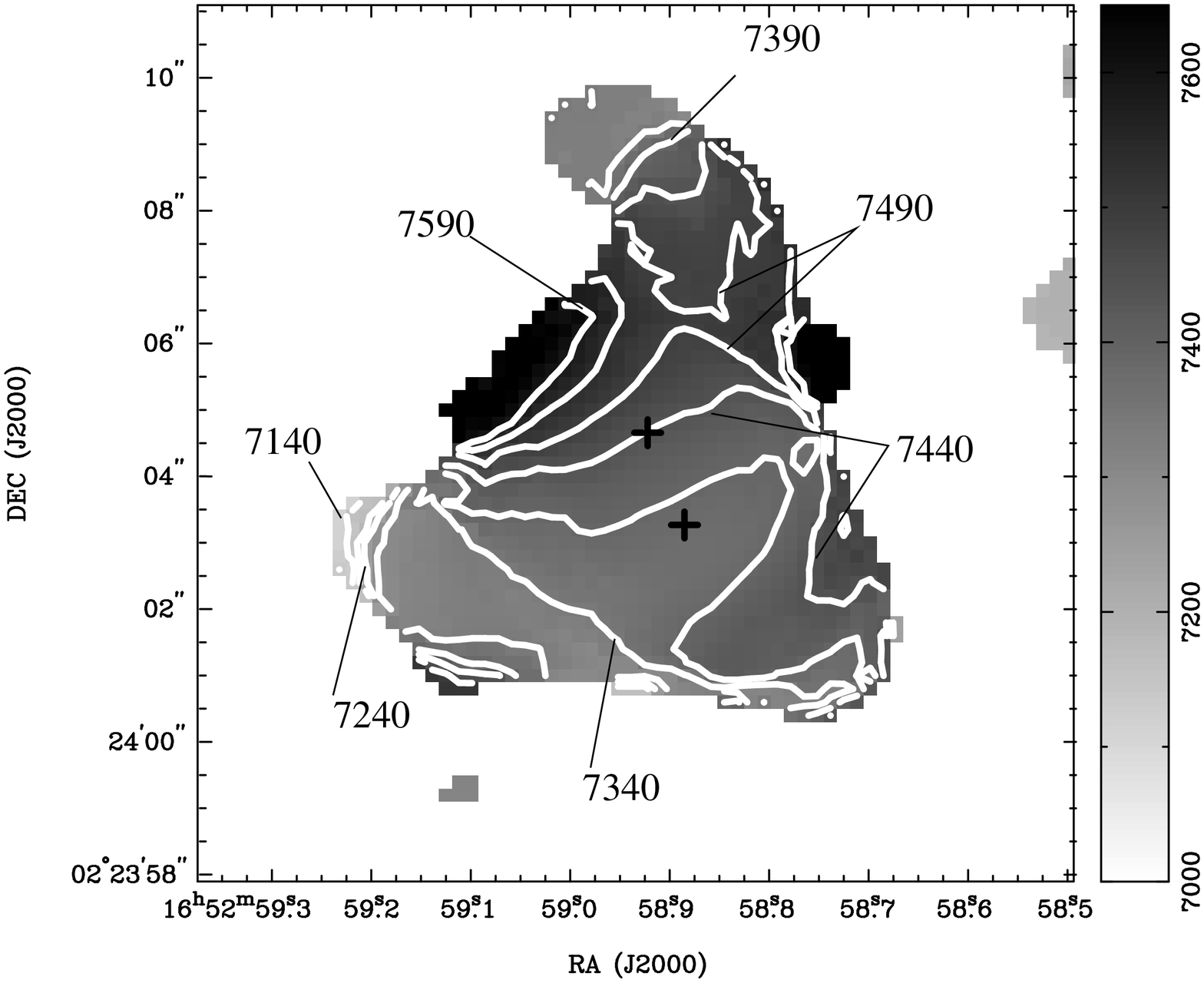}{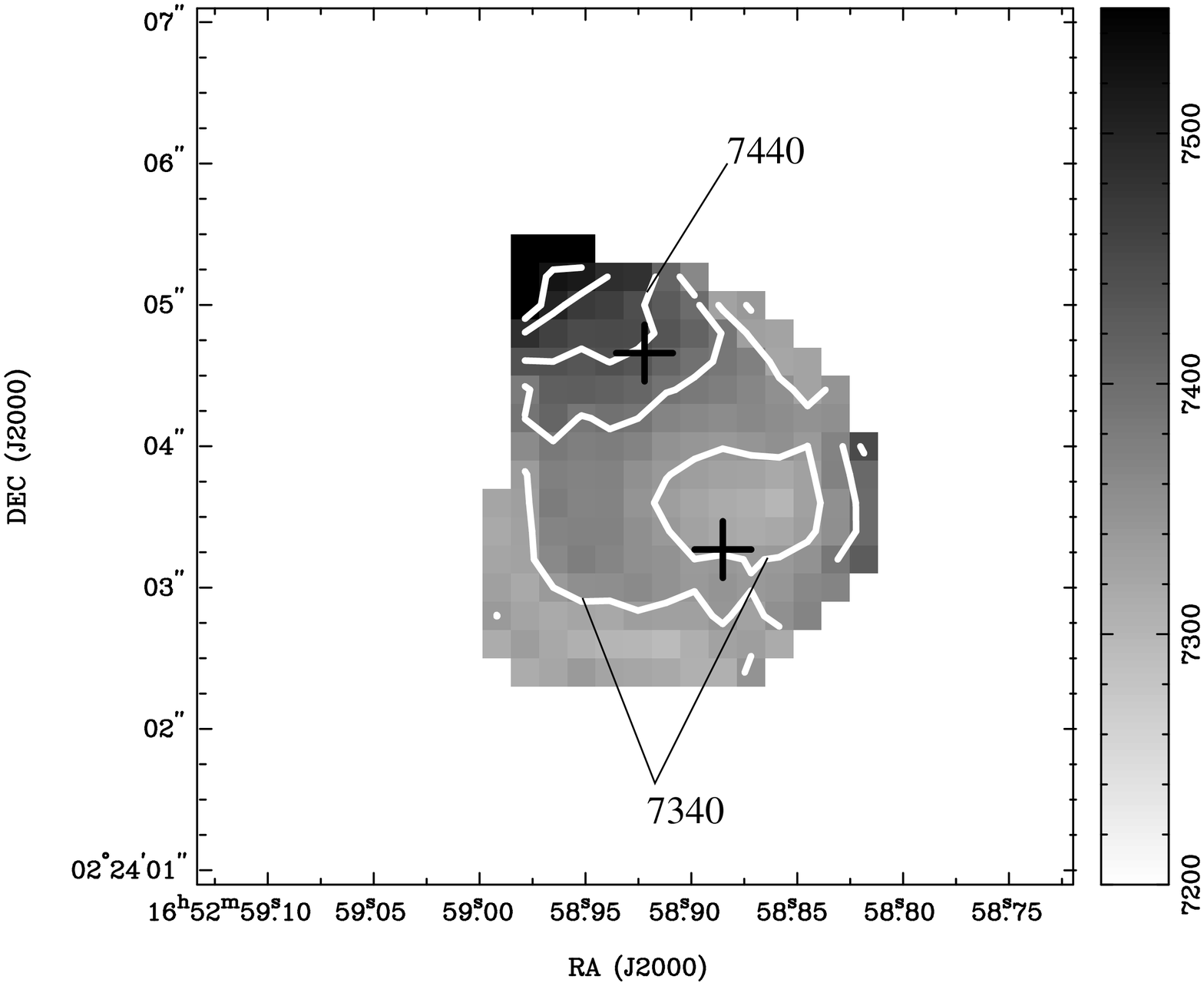}
  \caption{
    ($left$) The CO~(3--2) velocity distribution (moment 1) map from the 
    natural weighted map in the central $14''$ of NGC~6240.  The velocity 
    contours are spaced at 50~km~s$^{-1}$.  ($right$) same as $left$ but
    using uniform weighting.  The corresponding moment 0 maps are shown 
    in Figure~\ref{fig4}.  The units are in km~s$^{-1}$.
  }
  \label{fig5}
\end{figure}

The total CO~(3--2) FWZI linewidth of 1175~km~s$^{-1}$ is extremely large
(see Figure~\ref{fig3}).  The large linewidth, particularly in the medium
surrounding the two AGNs, suggests extremely turbulent gas.  
While it is possible that the bulk of the gas seen in 
Figure~\ref{fig5}~($left$) is dominated by turbulent gas stirred during the
massive collision, the northeast-southwest 
velocity gradient centered around the 
northern nucleus (N1) may be suggestive of a disk rotation.  
The higher resolution image in Figure~\ref{fig5}~($right$)
also shows this compact, rotation feature.
Rapid changes in velocity of $\Delta v \sim 100$~km~s$^{-1}$
are seen along the northern extended feature that appears to connect to this
rotating disk, possibly 
suggesting collimated gas streaming toward the direction close to the line 
of sight.  A steep velocity gradient like this is suggestive of 
gas inflow or outflow.  The velocity gradient in the vicinity 
of the southern nucleus (S1) is less obvious, possibly due to the dominance 
of local non-circular motion, or a face-on orientation of a rotating disk.
Although our current data does not have high enough angular resolution to 
distinguish the exact origin, if the southern nucleus has a face-on rotating 
disk, then the large shift in velocity of $\sim 250$~km~s$^{-1}$ that begins 
from the south of S1 and along the eastern extension may also 
suggest inflow or outflow activity.
The velocity distribution (Figure~\ref{fig5}~($right$)) 
shows that the systemic velocities of the two nuclei are offset 
by $\sim100$~km~s$^{-1}$.
The emission in WC is significant in the velocity range 7114 to 
7564~km~s$^{-1}$ (see Figure~\ref{fig3}), 
and does not show any significant evidence of a rotation 
in the moment map (Figure~\ref{fig6}).  
An overall velocity shift from the 
northern part to the southern part of WC is evident.

The position-velocity diagrams (PVD) sliced along the two AGNs (N1 and S1) and 
centered exactly halfway between the two AGNs are shown in  
Figure~\ref{fig7} for different resolution maps.  Figure~\ref{fig7}~($left$)
clearly shows that the linewidth near the two nuclei 
is extremely large, with emission slightly lopsided toward the 
redshifted velocities ($v>0$~km~s$^{-1}$). 
The PVD obtained with the uniformly weighted map resolves out the extended
emission in the velocity extrema, showing more details in the central 
$\pm 200$~km~s$^{-1}$ (Figure~\ref{fig7}~($middle$)).
Similar to Figure~\ref{fig7}~($left$), the peak of the emission is near 
the systemic velocity of 7339~km~s$^{-1}$.   
In contrast, the PVD obtained (Figure~\ref{fig7}~($right$)) 
with the highest angular 
resolution map completely resolves the central emission into two 
distinct peaks separated by $\sim70$~km~s$^{-1}$ and $\sim0\farcs2$~(100~pc).
This double peaked morphology in the PVD suggests overlapping molecular 
cloud complexes in the medium between the two AGNs.
The fact that these two peaks are nearly spatially coexistent, 
but separated in velocity further strengthens the argument that the gas 
kinematics in the central region is highly turbulent and non-circular.

\subsection{HCO$^+$(4--3) Emission}

The distribution of the HCO$^+$(4--3) emission (Figure~\ref{fig8}~($left$)) 
is highly compact, and it is concentrated in the central region with a 
northeast-southwest elongation of $4'' \times 2''$~($2 \times 1$~kpc), with 
an additional slight extension toward the southeast.  The peak of the 
HCO$^+$(4--3) emission is 
$\alpha(\rm J2000) = 16^{\rm h} 52^{\rm m} 58.9^{\rm s}$ 
and $\delta(\rm J2000) = 2^{\circ} 24' 03\farcs8$, 
and this is consistent with the peak of the CO~(3--2) to within the 
astrometric accuracy of the SMA.  
The integrated intensity is ($28\pm4$)~Jy~km~s$^{-1}$, recovering 
40\% of the total flux measured using a single dish telescope 
\citep[$78\pm16$~Jy~km~s$^{-1}$;][]{greve06}.
The peak brightness temperature observed with the
$\sim 2''$ beam is 0.5~K.
The highly compact and low S/N emission prevents us from constructing 
an interpretable velocity distribution map.
The HCO$^+$(1--0) emission obtained with a coarser synthesized beam
also peaks at a similar location, although the emission is more
extended in the northwest--southeast direction~\citep{nakanishi05}.

\subsection{$880~\micron$ Continuum Emission}

Similar to the CO~(3--2) and  HCO$^+$(4--3) emission, 
the $880\micron$ emission also peaks between the two nuclei, 
near the radio continuum feature S2. 
The $880~\micron$ peak is also consistent with the 1.3~mm peak
within the uncertainties of both measurements (Figure~\ref{fig1}).  
The coordinates of the $880\micron$ peak are 
$\alpha(\rm J2000) = 16^{\rm h} 52^{\rm m} 58.9^{\rm s}$  and
$\delta(\rm J2000) = 2^{\circ} 24' 03\farcs4$.  
The extension to the east is 
coincident with the similar extension seen in the CO~(3--2) map, 
but the continuum is much shorter.  The size of the $880~\micron$
emitting region is $6\farcs0 \times 3\farcs5$~($2.9 \times 1.7$~kpc).
The total flux detected from 
the SMA observations is ($105 \pm 6$)~mJy, 
and this is consistent with the true single dish continuum measurement of
96~mJy, which is calculated from the SCUBA $850\micron$ flux
\citep[][]{klass01} and subtracting the contribution of the CO~(3--2) 
emission using the formulae given in \citet{seaquist04}.
Assuming that all of the $880\micron$ emission is associated with dust, 
the derived dust mass in the central 2~kpc is 
$(4.7 \pm 1.4)\times 10^7$~M$_\odot$ 
using the formula described in Wilson et al. (2006 in prep).
This yields a gas-to-dust mass ratio of $(147\pm57)$ in the central 
2~kpc region (see \S3.1 for the gas mass).  
Although uncertainties in both mass calculations are large, 
this is in agreement with  
the values typically observed in the Galaxy.  The peak of the 
continuum suggests that the bulk of the dusty starburst activity occurs 
in this region. This is in contrast to previous optical/NIR imaging studies 
\citep[e.g.][]{pasquali04} where the visible starburst activity is 
concentrated near the two optical/X-ray nuclei, suggesting that the 
SFRs \citep[2.8 -- 412~M$_\odot$~yr$^{-1}$;][]{pasquali04} 
derived using these measurements are likely underestimated.  

\section{Analysis of the Physical Properties}
\subsection{Large Velocity Gradient Modeling}
Our new CO~(3--2) and HCO$^+$(4-3) data allow us to investigate 
the physical conditions of the gas by taking the 
line ratios with previously published CO~(1--0) and HCO$^+$(1--0) 
data \citep{nakanishi05}.  Past multi-line studies toward galaxies 
have been limited to line ratios obtained from 
single dish data \citep[e.g.][]{aalto95,petitpas00,yao03} or the 
ratios between the integrated intensities for nearby systems 
\citep[e.g.][]{matsushita99,mao00}.  Here we
present line ratios obtained in each velocity channel, 
and investigate the gas properties 
by comparing the ratios with the theoretical predictions obtained 
from Large Velocity Gradient (LVG) models \citep{goldreich74,takakuwa98}. 
We assume here that both CO and HCO$^+$ emission in the central kpc 
arise from the same molecular cloud. 
Although it is suggested that the filling factors of denser gas tracers may
be smaller than the commonly observed CO~(1--0) line \citep[e.g.][]{aalto95},
here we assume for simplicity 
that the (beam and volume) filling factors between 
CO~(3--2) and CO~(1--0), as well as HCO$^+$(4--3) and HCO$^+$(1--0) are 
similar.  We note that, although the HCO$^+$(4--3) emission recovers 
only 40\% of the total 
single dish flux (\S3.2), the main focus of this analysis is to investigate
the gas properties in the central $\sim 1''$ where the missing flux should
be negligible.  
By comparing the line ratios between CO~(3--2) and CO~(1--0), and also 
HCO$^+$(4--3) and HCO$^+$(1--0), these assumptions allow a 
direct comparison with LVG models that does not depend on the filling factor.

All data were convolved to the angular resolution of the CO~(1--0) image
($2\farcs2 \times 2\farcs1$; lowest angular 
resolution among the four data cubes), 
regridded to the same velocity resolution of 56~km~s$^{-1}$, and two 
line ratio cubes, CO~(3--2)/CO~(1--0) ($R_{\rm CO}$) and 
HCO$^+$(4--3)/HCO$^+$(1--0) ($R_{\rm HCO^+}$) were formed.  
The pixel-to-pixel correlation between the two ratio cubes were obtained 
and are shown as crosses in Figure~\ref{fig9}.  
In order to show the regions with significant 
detection in all four transitions, only the data above two sigma in each 
channel are plotted.  
In brightness temperature, the two sigma cutoffs are 
0.20~K, 0.80~K, 0.14~K, and 0.26~K, for CO~(3--2), CO~(1--0), 
HCO$^+$(4--3), and HCO$^+$(1--0) respectively.
Since the HCO$^+$(4--3) emission was 
only detected in the central -100 to 200~km~s$^{-1}$, only 
the densest gas in the central $< 2''$~(1~kpc) is traced using 
this method.

In order to understand the correlations seen in Figure~\ref{fig9} better, 
a series of LVG calculations were performed.  We adopt an LVG 
analysis that assumes a single cloud with spherical geometry, 
which requires three independent
parameters as inputs: the H$_2$ density
($n_{\rm H_2}$), kinetic temperature (T$_{\rm K}$), and the molecular
abundance per unit velocity gradient ($X/(dv/dr)$).   
The collision rates were obtained from \citet{flower85} for CO and
\citet{flower99} for HCO$^+$.
The quantity $X/(dv/dr)$, which is different for different 
species, is physically the most ambiguous parameter.  
In order to decrease the number of unknown parameters, we adopt 
a constraint to the relative abundance between CO and HCO$^+$
\citep[$X_{HCO^+} \sim 10^{-4} X_{CO}$;][]{bergin97} 
obtained from chemical evolution models.
By varying the CO abundance per unit velocity gradient from 
$X_{CO}/(dv/dr) = 10^{-4}$ to $10^{-9}$~(km~s$^{-1}$~pc$^{-1}$)$^{-1}$, 
$X_{CO}/(dv/dr) \sim 10^{-7}$~(km~s$^{-1}$~pc$^{-1}$)$^{-1}$ 
provides the most sensible explanation of the data.
Changing $X_{CO}/(dv/dr)$ to 10$^{-6}$~(km~s$^{-1}$~pc$^{-1}$)$^{-1}$ 
will shift the temperature 
(vertical) contours leftward (see Figure~\ref{fig10}), 
yielding most of the molecular gas with temperatures in the range 100 -- 200~K.
This is higher than the cold and warm dust temperatures predicted from
a three component model by \citet{armus06} where they find cold (27~K),
warm (81~K) and hot (680~K) dust components to coexist over the entire NGC~6240
system.  Using $X_{CO}/(dv/dr) = 10^{-8}$~(km~s$^{-1}$~pc$^{-1}$)$^{-1}$ 
will shift the temperature contours rightward, yielding 
most of the data points in T $< 40$~K.  This is lower 
than the values predicted in the nearby starburst NGC~253 \citep{jackson95},
but consistent with the cold component dust in NGC~6240 suggested 
by \citet{armus06}, assuming that the gas and dust are in thermal equilibrium.

Assuming that the gas temperature of 40 -- 100~K  is a reasonable 
representation of the molecular clouds in NGC~6240, then 
$X_{CO}/(dv/dr) \sim 10^{-7}$~(km~s$^{-1}$~pc$^{-1}$)$^{-1}$ 
provides the most sensible abundance that explains the data in all channels
with significant emission (Figure~\ref{fig9}).
A tight constraint to the H$_2$ density of 
$n_{\rm H_2} = 10^{5.0 - 5.4}$~cm$^{-3}$ is obtained.
The derived opacities range from $\tau = 0.01$ -- 0.2 for the 
CO~(1--0) line, $\tau = 0.2$ -- 2 for the CO~(3--2) line,
and $\tau = 0.5$ -- 1 for the HCO$^+$(4--3) line.  The model predicts that 
the relative population of the HCO$^+$(1--0) line be inverted, thus giving
negative opacities and excitation temperatures.
The excitation temperatures (T$_{ex}$) derived from these analyses 
show that both CO lines are nearly thermalized 
(i.e. T$_{\rm ex, CO(3-2)}$~$\sim$~T$_{\rm ex, CO(1-0)}$).

Once these physical quantities are derived, it is now possible to 
constrain the beam filling factor of the CO emitting molecular gas.  
The observed line intensities are diluted by the $\sim 2''$~(1~kpc) beam, 
and thus the true flux is obtained by scaling the observed flux by
the reciprocal of the filling factor.
Figure~\ref{fig11} and \ref{fig12} 
show the correlation between T$_{\rm CO~(3-2)}$ and 
T$_{\rm CO~(1-0)}$ in each 21~km~s$^{-1}$ 
channel where the emission of each line is significant   
(i.e. the central -133 to 187~km~s$^{-1}$).
The filling factors of the CO lines, which are assumed to be the same
between CO~(3--2) and CO~(1--0) as well as in all velocity channels,
are varied from 1 (no scaling; Figure~\ref{fig11}) to 0.2 
(multiply the observed brightness temperatures by 5; Figure~\ref{fig12}). 
The lower limit to the filling factor of 0.2 is obtained by adjusting 
the filling factor until the highest observed points reach  
$n_{\rm H_2} \lesssim 10^{5.4}$~cm$^{-3}$, a constraint
given by the R$_{\rm CO}$ -- R$_{\rm HCO^+}$ plot in Figure~\ref{fig9}.
A CO filling factor of unity gives the bulk of the gas in the temperature 
range of 40 -- 200~K and density of $n_{\rm H_2} \sim 10^{4.0}$~cm$^{-3}$.
In this case, while the range of temperature is consistent with our
earlier analysis (see above), the H$_2$ density is an order of magnitude 
lower.  A CO filling factor of 0.2 also yields a temperature
range of 20 -- 100~K, and the range of H$_2$ density is also comparable to our 
earlier analysis.  These results suggest that the true 
CO filling factor is between 0.2 to 1, but probably closer to 0.2. 
A low value of the filling factor would be inconsistent with the 
previous studies by \citet{downes98} 
where they suggest that the diffuse inter-clump medium may dominate the 
CO emission in ULIRGs, yielding a filling factor of unity. 

An estimate of the CO beam filling factor allows us 
to derive the volume filling factor of the CO emitting gas 
\citep[see Equation 3 of][]{sakamoto94}.
Assuming that the CO~(3--2) emission arises from GMCs of order 50~pc 
each, and that the CO~(3--2) emission region along the line of sight 
extends 4~kpc, the range of possible volume filling factor 
is 0.004 -- 0.05 for filling factors 0.2 -- 1.  Changing the size of the 
GMC to 25~pc (10~pc) gives 0.002 -- 0.03 (0.001 -- 0.01) 
for the volume filling factor.
Using the average  density estimate from our LVG analysis 
(i.e. $n_{\rm H_2} \sim 10^{5.0}$~cm$^{-3}$), 
the derived molecular gas mass is 
then $10^{9 - 10}$~M$_{\odot}$ in the central 0.5~kpc (r=0.25~kpc) 
region. Our mass estimate is consistent with $6.9 \times 10^9$~M$_{\odot}$
derived using X=0.8~M$_\odot$~(K~km~s$^{-1}$)$^{-1}$~pc$^2$ (see \S3.1), 
and with $(1-4) \times 10^9$~M$_\odot$ derived in the central 0.5~kpc
using CO~(2--1) line emission \citep{tacconi99}.
Assuming that the average density in the larger spatial extent is
also $n_{\rm H_2} = 10^{5.0}$~cm$^{-3}$  gives
$\sim 10^{10}$~M$_{\odot}$ in the central 1~kpc (r=0.5~kpc), and 
$\sim 10^{11}$~M$_{\odot}$ in the central 2~kpc (r=1~kpc) region.  The 
mass estimate given in a larger region (i.e. 2~kpc) is probably 
an overestimate since the molecular gas density in the periphery of the 
nucleus is likely lower than the central 0.5~kpc.  
Since the density derived from the LVG analysis
only accounts for the densest gas in the central region, 
a proper mass estimate must consider the density variation 
of the cloud in the entire 1175~km~s$^{-1}$ velocity width, 
which is not possible here because of the varying mass sensitivities among 
the four data cubes.

The LVG calculation also provides an estimate of the H$_2$ column density from
the relation $3.0\times 10^{18}$ 
(X$_{\rm CO}$/(dv/dr)) $n_{\rm H_2} = $ N$_{\rm CO}$/dv.
By adopting 
$X_{CO}/(dv/dr) = 10^{-7}$~(km~s$^{-1}$~pc$^{-1}$)$^{-1}$ and
$n_{\rm H_2} = 10^{5.0}$~cm$^{-3}$, we find 
N$_{\rm CO}$/dv = $3.0 \times 10^{16}$~cm$^{-2}$[km~s$^{-1}$]$^{-1}$.  Using
the standard abundance observed in galactic sources
\citep[X$_{\rm CO}$/X$_{\rm H_2}=10^{-4}$;][]{blake87} and 
dv~=~FWZI~=~1175~km~s$^{-1}$, we obtain
N$_{\rm H_2} = 4.0 \times 10^{23}$~cm$^{-2}$.  
This is in excellent agreement with our previous
column density estimate derived using the peak CO~(3--2) flux.

The peak brightness temperatures obtained in the 
CO~(3--2) and $880\micron$ observations (see \S3.1 and \S3.3) allow 
a confirmation of the properties derived above.  In general,
the brightness temperature and the true excitation temperature are 
related by T$_{B}$~=~$f$~T$_{ex}$~($1-e^{-\tau}$), where T$_{B}$, $f$, T$_{ex}$
and $\tau$ are the observed brightness temperature, beam filling factor, 
excitation temperature, and opacity respectively.  Assuming the observed
molecular clouds and the CO excitation are in thermal equilibrium 
coupled with dust (i.e. T$_{ex} \sim$ T$_{K} \sim $ T$_{dust} \sim $ 49~K), 
and using the range of 
filling factors of $f =0.2$ -- 1 (see above), 
we obtain $\tau = 0.2$ -- 1.2, which
is consistent with the opacities derived above.  This is further evidence 
that the observed molecular clouds are moderately 
optically thin, and that they are close to thermal equilibrium. 
A similar analysis is performed on the dust continuum, and we derive
the dust opacity $\tau_{dust} = 0.003$ -- 0.01 using the same 
filling factors as CO, suggesting that
the cold dust traced in 880$\micron$ is indeed optically thin.

In summary, the important result obtained from these quantitative analyses 
is that the peak of the molecular gas in the central
1~kpc of NGC~6240 is warm, dense, and moderately optically thin.  
The derived temperatures and densities of NGC~6240 are consistent with 
the high values found in the nearby starburst galaxies M82 and NGC~253 
\citep{wild92,jackson95,seaquist00}.  The low opacities 
predicted from these analyses are somewhat consistent with earlier results
that investigate the opacities in M82 \citep[$\tau = 0.5 - 4.5$;][]{mao00}, 
but our new results predict much lower opacities than previously suggested
in ULIRGs \citep[$\tau = 3 - 10$;][]{downes98}.  
The low opacities found in the central region of NGC~6240 are possibly
due to large velocity gradients caused by extremely turbulent gas,  
allowing emission in the core of the molecular cloud to escape efficiently. 
In addition, 
our predicted peak densities ($n_{\rm H_2} = 10^{5.0 - 5.4}$~cm$^{-3}$) 
are much higher than the range of values found
in ULIRGs using low $J$ transition CO emission 
\citep[$n_{\rm H_2} = 10^{2.3 - 4.3}$~cm$^{-3}$;][]{downes98}.
The disagreement
is likely due to the difference in the excitation conditions of the observed
molecular gas tracers; i.e. \citet{downes98} used cold (T~$\sim10$~K) 
and low density ($n_{crit} \sim 10^3$~cm$^{-3}$) 
tracers whereas our new observations
trace much warmer (T~$\sim30$~K) and 
denser ($n_{crit} \sim 10^4$~cm$^{-3}$ for CO~(3--2)) molecular gas.
This study, therefore, demonstrates the importance of using high density 
submm line tracers such as CO~(3--2) and HCO$^+$ emission to study the 
physical conditions in the nuclear ISM in ULIRGs.

\subsection{The Gas to Dust Mass Ratio}

The gas-to-dust mass ratio (i.e. M$_{\rm H_2}$/M$_d$) provides an important
measure of the relative abundance between gas and metallicity.  An average
M$_{\rm H_2}$/M$_d$ over the entire galaxy is often derived in single 
dish work, where it is found that M$_{\rm H_2}$/M$_d \sim 100$ for Galactic
sources \citep{hildebrand83}, M$_{\rm H_2}$/M$_d \sim 200$ -- 300 for 
local LIRGs/ULIRGs \citep{contini03, yao03, seaquist04}, and 
M$_{\rm H_2}$/M$_d$ = 15 -- 231 in high-$z$ sources \citep{solomon05}.
Wilson et al. (2006 in prep) finds M$_{\rm H_2}$/M$_d = 357 \pm 95$ from 
a sample of 13 LIRGs/ULIRGs observed in high resolution with the SMA.  
It is found that the global gas-to-dust 
mass ratio in NGC~6240 is M$_{\rm H_2}$/M$_d = (147 \pm 57)$~(see \S3.3).

Our new high angular resolution CO~(3--2) and $880~\micron$ maps 
allow us to study the spatial variation of M$_{\rm H_2}$/M$_d$.
We have converted the CO~(3--2) and $880\micron$ maps to  
M$_{\rm H_2}$ and M$_{\rm d}$ maps.
The gas-to-dust mass ratio map is shown in Figure~\ref{fig13}.  
The peak of M$_{\rm H_2}$/M$_d$ occurs between the two AGNs, and near S2, 
suggesting an over abundance of dense gas relative to dust in 
these regions, assuming that most of the $880\micron$ emission is tracing 
dust.  The high ratios seen north of the northern AGN are due to low 
dust masses at the edge of the 3$\sigma$ cutoff.  The peak 
ratio is $\sim 150$.  We note that the ratio map obtained here is only
available in the central kpc where the dust emission is 
significant ($> 3 \sigma$).  The CO~(3--2) emission is much more 
extended than the dust emission and hence the average M$_{\rm H_2}$/M$_d$
in Figure~\ref{fig13} 
is lower than the global M$_{\rm H_2}$/M$_d$ ($147\pm57$).
Higher ratios are seen in the north-south direction 
(i.e. M$_{\rm H_2}$/M$_d \sim$ 100 -- 160) along the two AGNs 
than the east-west direction (i.e. M$_{\rm H_2}$/M$_d \sim$ 30 -- 100).  
The steep decline in the wings of dust emission
may suggest substantial free-free emission associated with the
starburst winds, or a local gradient in the metallicity.

\subsection{The FIR to CO~(3--2) Luminosity Ratio}

It is empirically known that the ratio between FIR and CO~(1--0) 
luminosities is higher for ULIRGs than for normal spiral 
galaxies \citep{solomon97}.  \citet{solomon97} attribute this 
result to the presence of compact, optically thick dust emission, 
which in turn yields higher dust temperatures than in the 
optically thin limit.  However, these measurements are performed using 
the CO~(1--0) transition, and it is found that this value is different when 
the ratio between FIR and CO~(3--2) is taken 
\citep[i.e. compare log~L$_{\rm FIR}$/log~L$_{\rm CO(3-2)} = 1.43 \pm 0.08$
with log~L$_{\rm FIR}$/log~L$_{\rm CO(1-0)} = 1.72 \pm 0.21$;][]{yao03}.
The FIR--to--CO~(3--2) luminosity ratio appears to be even lower  
in high-$z$ sources where the FIR luminosity
is typically an order of magnitude larger 
(i.e. L$_{FIR} \sim 10^{13}$~L$_{\odot}$). From the high-$z$ sources with
a robust CO~(3--2) detection, the ratio is found to be
log~L$_{\rm FIR}$/log~L$_{\rm CO(3-2)} = 1.23 \pm 0.03$ \citep{solomon05,iono06}. 
Using the CO~(3--2) integrated intensity from our full resolution map, and the
IRAS measurement for the FIR luminosity, 
the ratio for NGC~6240 is found to be 
log~L$_{\rm FIR}$/log~L$_{\rm CO(3-2)}$ = 1.16. 

NGC~6240 thus contains much more molecular gas relative to the 
intensity of starburst activity than 
similarly FIR bright systems observed in the local universe.
The ratio is even lower than the values derived in high-$z$ sources.  
The overwhelming amount of molecular gas
available in the central region of NGC~6240 suggests that there is enough
gas to support star formation for at least another $\sim 10^8$~years, 
assuming that all of the gas will eventually be converted to stars with 
a constant star formation rate of 60~M$_\odot$~yr$^{-1}$.  
The wealth of molecular gas may also suggest that
the star formation in NGC~6240 will increase to a level comparable to that seen
in Arp~220.  

\section{Discussion}

\subsection{The Nature of the Western Molecular Complex (WC)}


Recent galaxy-galaxy collision simulations \citep[e.g.][]{hopkins05} 
that trace the evolution of the central AGN have shown that significant 
outflow activity may be observable in high $J$ CO transitions up to 
$J=6-5$ \citep{narayanan06}. 
While a massive ($3.5 \times 10^8$~M$_\odot$) off-nuclear molecular complex 
such as WC is often explained by starburst superwinds \citep{heckman90}, 
we investigate here the possibility that the origin of WC 
is gas entrained in outflow 
from the AGN of one of the progenitor galaxies.  
From an estimate of the cloud diameter of $D=2.7$~kpc from the CO~(3--2) map
and the velocity dispersion of $\sigma_v = 400 \pm 100$~km~s$^{-1}$ estimated
from the FWHM of the CO~(3--2) line in WC, we find a virial mass of 
M$_{vir} = 3.0 \times 10^{11}$~M$_{\odot}$.  The error in this virial mass
estimate could be as large as a factor of (2--3)  due to 
large uncertainties in the cloud size and velocity dispersion.
By assuming the Galactic dynamical mass (M$_{dyn}$) to M$_{\rm H_2}$ ratio of 
136 \citep{nakanishi06}, we estimate a total mass of 
M$_{tot} = 4.8 \times 10^{10}$~M$_\odot$ for WC.  
Although the uncertainties in these measurements are large, 
the derived virial mass is an order of magnitude larger than the 
total mass of WC.  This suggests that WC is not 
bound by its own gravitational force and will likely disperse within 
$D/\sigma_v \sim 7 \times 10^6$~years. 
The systemic velocity of WC is $v_{sys} = 7449$~km~s$^{-1}$,
which yields a relative velocity 
$v_{\rm WC} \sim 210$~km~s$^{-1}$ assuming that WC is
ejected from the northern nucleus.  Using a projected distance between 
the northern nucleus and WC of 4.4~kpc, the timescale for WC to traverse to
the current location is $\sim 2 \times 10^7$~years.  

Although all of these timescales contain significant uncertainties, 
the short dispersion time ($7 \times 10^6$~years) 
suggests that it is unlikely that the
cloud would have survived the estimated transversal time of 
$2 \times 10^7$ years to reach the current location.
Simulations conducted by \citet{narayanan06} predict that the total gas mass
of an ejected cloud could be as large as $\sim 10^8$~M$_\odot$ with a
cloud dispersion time of $t \sim 2.9 \times 10^8$~years \citep{narayanan06}.
While the CO~(3--2) derived gas mass is consistent with the 
ejected gas mass predicted from simulations, the derived dispersion timescale 
is a factor of 40 shorter.  These results may suggest that the
AGN outflow scenario is less appealing to explain WC, and other scenarios 
such as molecular outflow from starburst superwinds 
may be more appealing to explain the large and extended molecular gas.
The starburst outflow scenario to explain the western extension of 
radio continuum emission in NGC~6240 was suggested by \citet[][]{colbert94}.

Possible scenarios to explain WC are not limited to starburst superwinds or
to AGN outflows
and both of the following are also possibilities; (1) molecular gas 
ejected from the violent arm-arm collision, (2) molecular gas complex 
formed locally as a result of gravitational instability.
Ejection of molecular gas is possible as arm-arm collision could
occur in each pericentric passage.  However, ejecting a molecular gas mass of
order $10^8$~M$_{\odot}$ and diameter 1 -- 2~kpc from the potential
of a tidal arm may require a significant amount of energy.
While it is possible that gravitational collapse could form a CO cloud 
from an ambient \ion{H}{1} cloud, it is already demonstrated that the
virial mass is an order of magnitude larger than the estimated 
total mass of WC.
Hence these two scenarios are less appealing to explain the origin of 
WC than the starburst superwind or the AGN outflow scenarios.
Our current data, however, is not sensitive enough to 
address the exact origin of WC, and future observations with higher 
angular resolution and sensitivity should provide us further understandings 
to the possibilities of these scenarios.


\subsection{The Central 1~kpc of NGC~6240}

Our new analysis presented in \S4.1 shows that the molecular gas in the
central 1~kpc of NGC~6240 is warm, dense and moderately 
optically thin.  These results suggest that the excitation
conditions of molecular gas surrounding the two AGNs is similar to the 
nuclear regions of nearby starbursting galaxies.  The low opacity is likely
caused by highly turbulent gas, which is also supported by the
dominance of non-circular 
kinematics entangled with possible disk-like rotation around the northern 
nucleus (see \S3.1.2).  The southern nucleus shows little evidence of
disk-like rotation possibly due to the dominance of non-circular motion, or
due to a face-on orientation of a rotating disk.  Although the projected 
nuclear separation is different, the double nucleus configuration is 
similar to that seen in Arp~220 where a merger of two counter-rotating disks 
is suggested from high resolution CO~(2--1) observations \citep{sakamoto99}.
Higher angular resolution observations should provide further insight to 
these possibilities.

In addition to the large amount of dense molecular gas, 
there is significant multi-wavelength evidence that  
the central 1~kpc also harbors substantial amounts of dust.  
While we caution here that emission traced in 
$880\micron$ and 1.3~mm \citep{tacconi99} are possibly more dominated 
by non-thermal emission, the ubiquitous presence of the optically thin 
dust emission in the medium between the two AGNs is a significant evidence 
of cold dust.  
Further, from comparison between star formation activity 
traced in optical and the AGN activity traced in X-ray, \citet{gerssen04} 
have shown that the two nuclear star forming regions in both progenitor disks 
are not coincident with the AGN positions but are offset by $\sim 0\farcs5$,  
suggesting that foreground dust extinction in the central region is 
significant. 
The dusty central 1~kpc also shows significant evidence of a coexistence 
of hot and shocked gas as seen in detections of 
maser emission \citep{hagiwara03} and H$_2$ emission \citep{ohyama00}. 
Both of these are detected near S1 and S2, providing evidence of shocked gas 
associated with either the AGN jet or the nuclear starburst activity.  
Summarizing the multi-wavelength properties, the central 1~kpc of NGC~6240 
harbors two AGNs surrounded 
by a wealth of warm and dense molecular gas, as well as 
hot (T~=~$10^{(3-7)}$~K) gas and a large amount of dust, and the 
region is undergoing a high rate of star formation activity.

\section{Summary}

We present high resolution CO~(3--2), HCO$^+$(4--3) and $880\micron$ 
imaging of the luminous infrared galaxy NGC~6240 obtained at the SMA.  
We provide our main findings below;
\begin{enumerate}
\item The distribution of CO~(3--2) is extended on a 4~kpc scale, but the
  HCO$^+$(4--3) emission is more compact and concentrated in the central
  1~kpc between the two nuclear AGNs.  
  The CO~(3--2),  HCO$^+$(4--3), and $880\micron$ emission all 
  peak between the two AGNs where the H$_2$ emission was detected previously.  

\item The kinematical information 
  provided by the CO~(3--2) emission shows a rotating disk-like feature 
  centered around the northern AGN, but the kinematics of gas between the two 
  nuclei is extremely turbulent. 
\item Our LVG analysis shows that the 
  gas near the peak of the continuum emission is warm (T = 20 -- 100~K), 
  dense ($n_{\rm H_2} = 10^{5.0-5.4}$~cm$^{-3}$) and moderately 
  optically thin ($\tau = 0.2$ -- 2) in the central 1~kpc of NGC~6240.
\item We propose that the isolated molecular emission with 
  $3.5 \times 10^8$~M$_{\odot}$ seen west of the main CO component 
  may be associated with outflows from starburst superwinds, 
  but the gas outflow scenario from one of the central AGN is not 
  completely ruled out.  
\end{enumerate}

\acknowledgements
We thank Kouichiro Nakanishi, Sachiko Okumura and Naomasa Nakai for providing
their published CO~(1--0) and HCO+(1--0) data cubes.  
We also thank Linda Tacconi for 
kindly providing us with the CO~(2--1) data cube.  
The Submillimeter 
Array is a joint project between the Smithsonian Astrophysical Observatory 
and the Academia Sinica Institute of Astronomy and Astrophysics, and is 
funded by the Smithsonian Institution and the Academia Sinica.

\clearpage

\begin{deluxetable}{llcc}
\tabletypesize{\scriptsize}
\tablewidth{0pt}
\tablecaption{Derived Properties}
\tablehead{
\colhead{Observation} & \colhead{Property} & \colhead{Value}
} 
\startdata
CO~(3--2) \\
& $v_{sys}$~(km~s$^{-1}$) & $7449\pm25$\\
& FWZI~(km~s$^{-1}$) & $1175\pm25$\\
& S$_{\nu}$d$\nu$~(Jy~km~s$^{-1}$) & $2834\pm 20$\\
& peak R.A. & $16^{\rm h} 52^{\rm m} 58.9^{\rm s}$ \\
& peak Decl. & $2^{\circ} 24' 03\farcs7$ \\
& peak T$_B$~(K) & 8.6 \\
HCO$^+$(4--3) \\
& $v_{sys}$~(km~s$^{-1}$) & $7339 \pm50$\\
& FWZI\tablenotemark{1}~(km~s$^{-1}$) & $300\pm50$\\
& S$_{\nu}$d$\nu$\tablenotemark{1}~(Jy~km~s$^{-1}$) & $28 \pm 4$\\
& peak R.A. & $16^{\rm h} 52^{\rm m} 58.9^{\rm s}$ \\
& peak Decl. & $2^{\circ} 24' 03\farcs8$ \\
& peak T$_B$~(K) & 0.5 \\
$880\micron$ continuum \\
& S$_{880}$~(mJy) & $105 \pm 6$\\
& peak R.A. & $16^{\rm h} 52^{\rm m} 58.9^{\rm s}$ \\
& peak Decl. & $2^{\circ} 24' 03\farcs4$ \\
& peak T$_B$~(K) & 0.1 \\
\enddata
\tablenotetext{1}{The FWZI and S$_{\nu}$d$\nu$ of the HCO$^+$(4--3)
emission may be underestimated since the emission occurs near the 
edge of the bandpass (see text).}
\end{deluxetable}

\clearpage

\begin{figure}
  \plotone{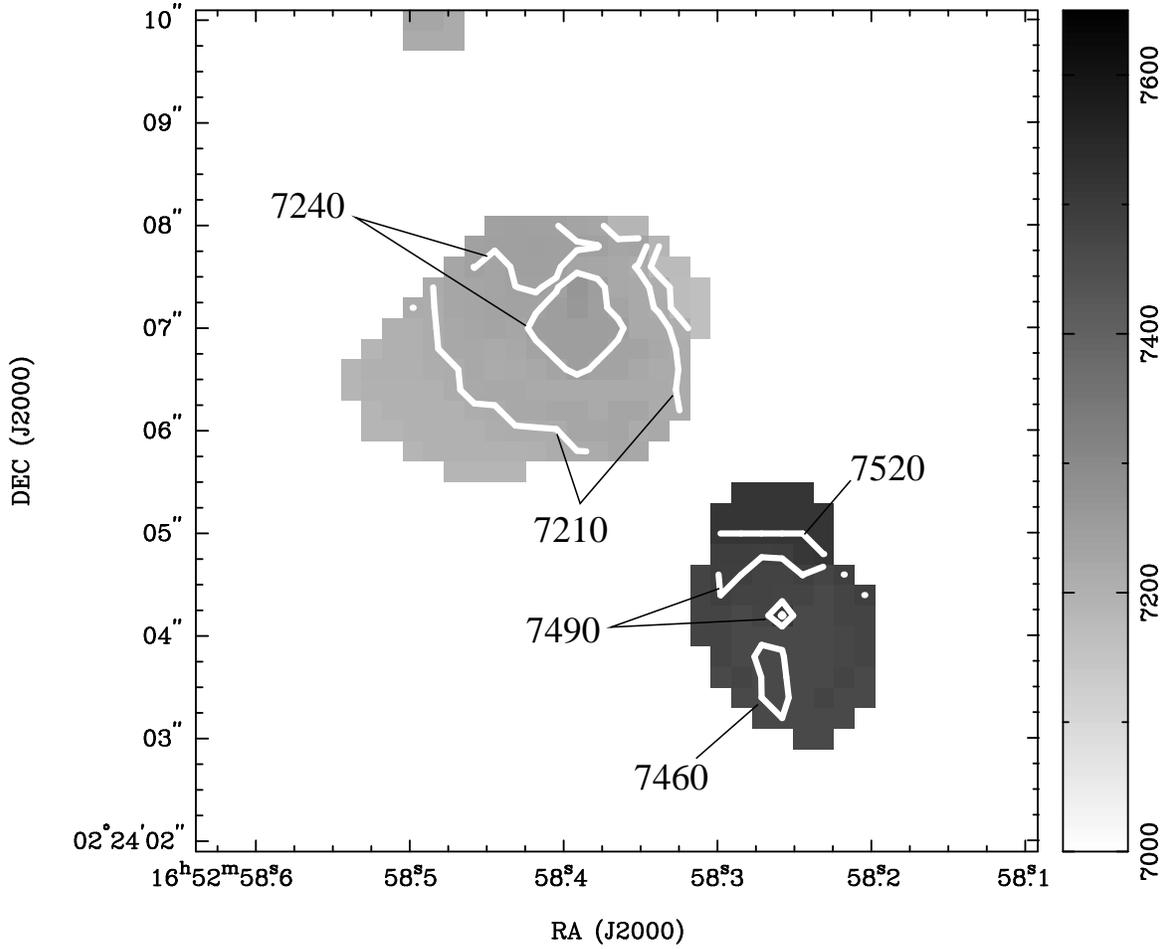}
  \caption{
    The CO~(3--2) velocity distribution of the western complex (WC) alone.  
    The velocity contours are spaced at 30~km~s$^{-1}$.
    The units are in km~s$^{-1}$.
}
  \label{fig6}
\end{figure}

\begin{figure}
  \plotone{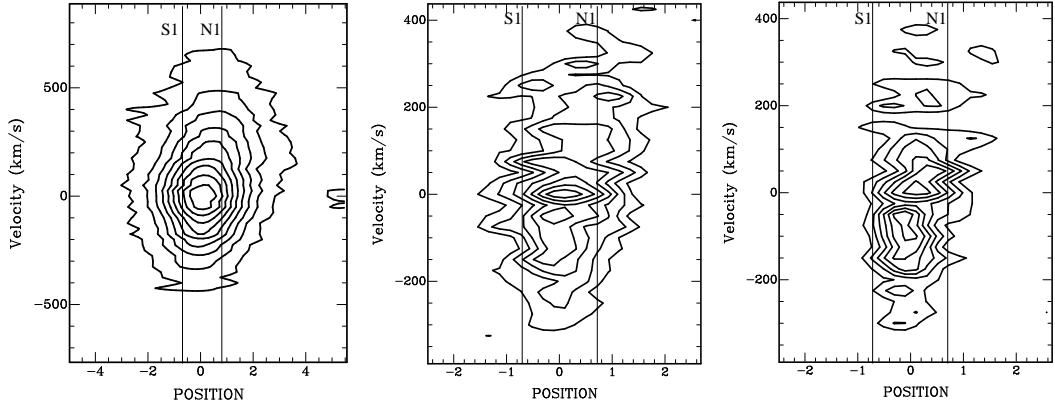}
  \caption{
    The position velocity diagram for ($left$) the $2\farcs0 \times 1\farcs8$
    resolution map (natural weighting).   The x-axis shows the 
    offset position from the center in arcseconds, and the y-axis 
    shows the velocity in km~s$^{-1}$.  The contours represent 
    10,20,30,40,50,60,70,80,90\% of the peak.  
    These PVDs are made by slicing the cube along the two AGNs (N1 and S1) and 
    centered exactly halfway between the two AGNs.   
    The location of the two AGNs are shown as vertical lines.
    The linewidth near the two nuclei 
    is extremely large, with emission slightly lopsided toward the redshifted
    velocity.  ($middle$) Similar to $left$ but for the  
    $1\farcs1 \times 0\farcs8$ resolution map (uniform weighting).  
    The contours represent the 1,5,10,20,40,60,80\% of the peak. 
    and ($right$) $0\farcs9 \times 0\farcs5$ resolution map 
    (using the longest baselines only).  The contours represent the 
    15,30,45,60,75,90\% of the peak.
    The nuclear region is resolved into two distinct components in the 
    highest resolution map.
  }
  \label{fig7}
\end{figure}

\begin{figure}
  \plottwo{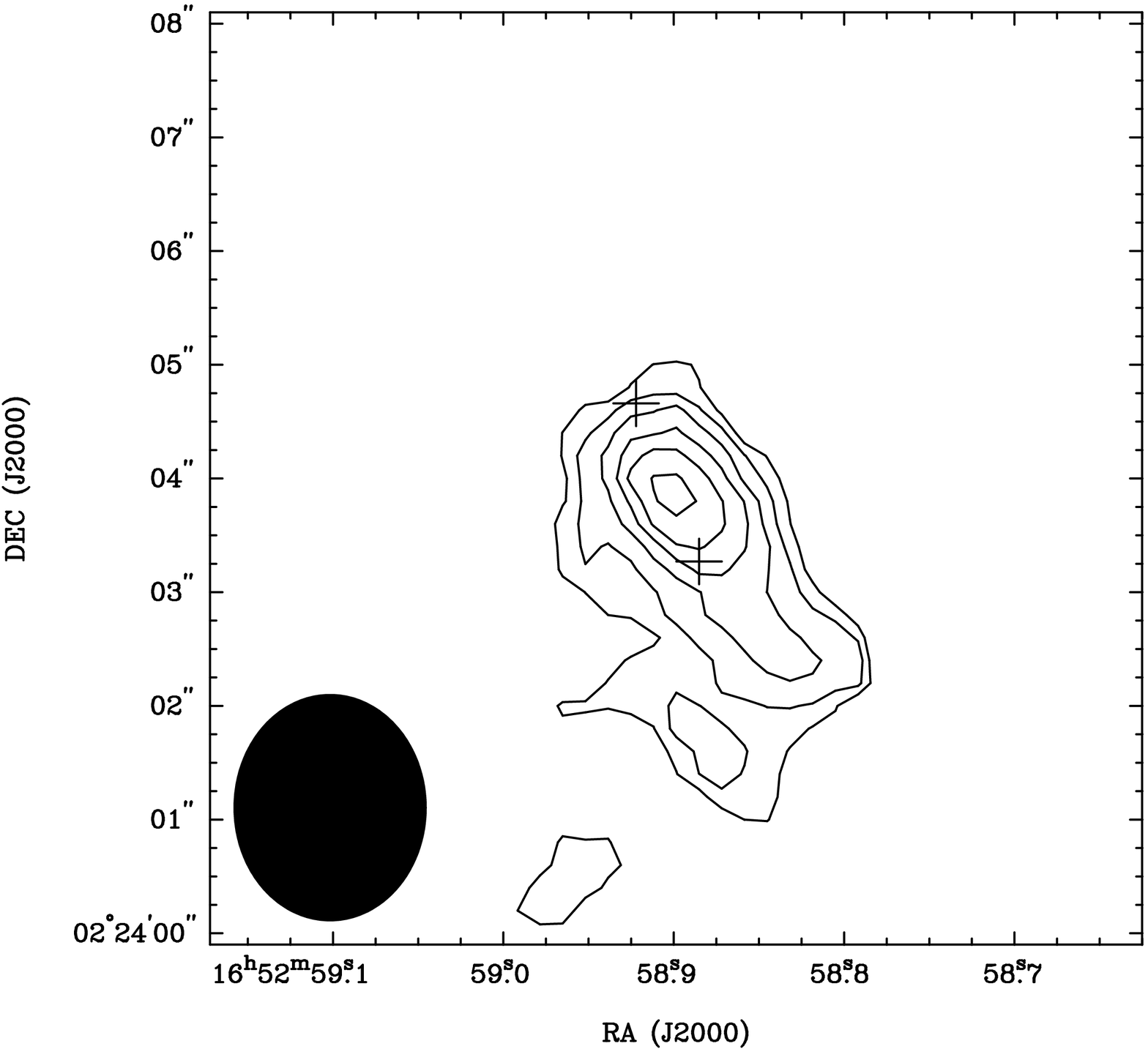}{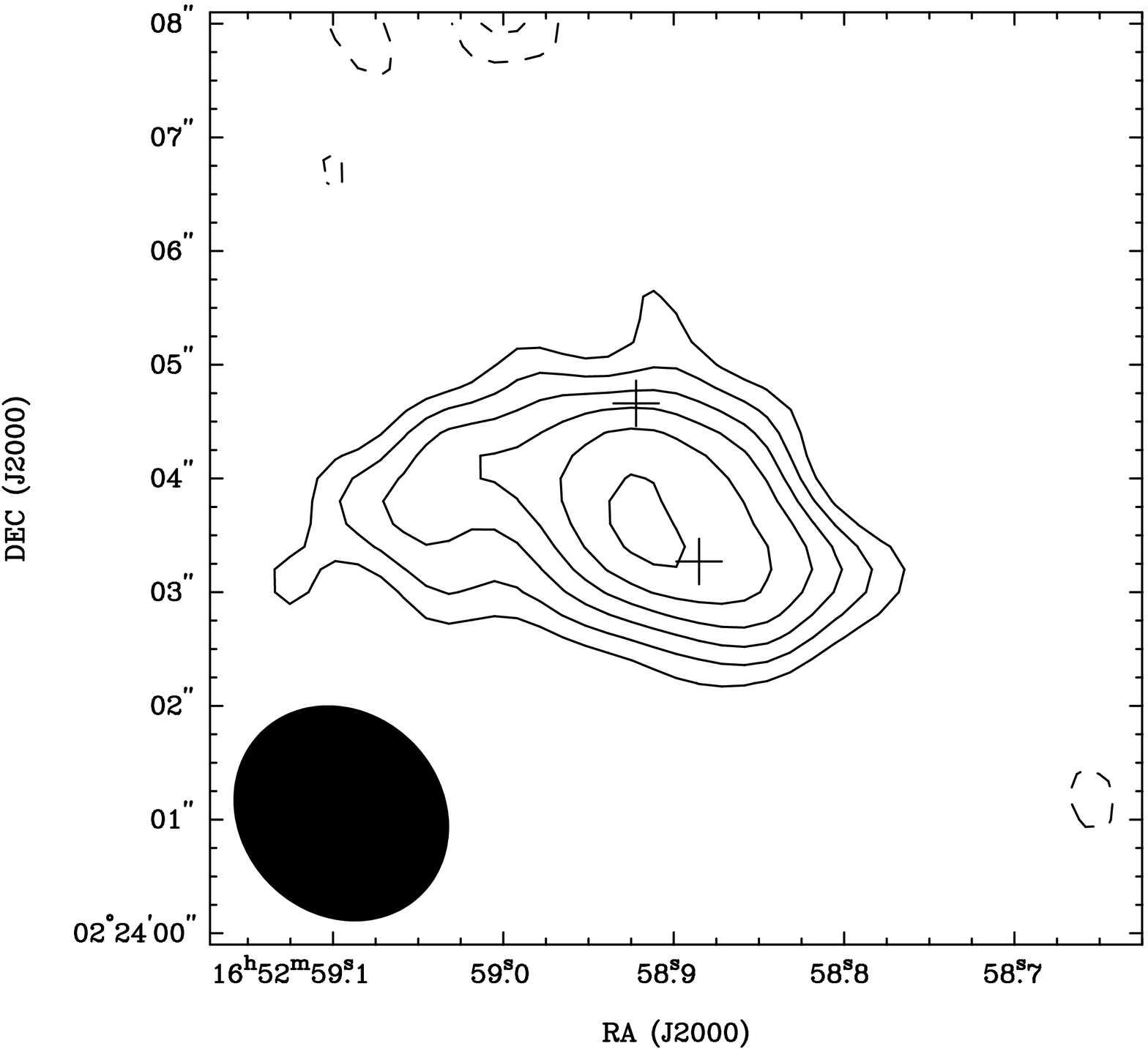}
  \caption{
    ($left$) The integrated intensity map of the HCO$^+$(4--3) emission.
    The contours are 5~Jy~km~s$^{-1}$~($\times$ 3,4,5,6,7,8).
    ($right$) The 880$\micron$ emission map.  The contours are 
    6~mJy~($\times$ -4,-3,3,4,5,6,7).  Both of these maps are made using
    natural weighting, and 
    the symbols in both maps are the same as in Figure~\ref{fig4}.
}
  \label{fig8}
\end{figure}

\begin{figure}
  \plotone{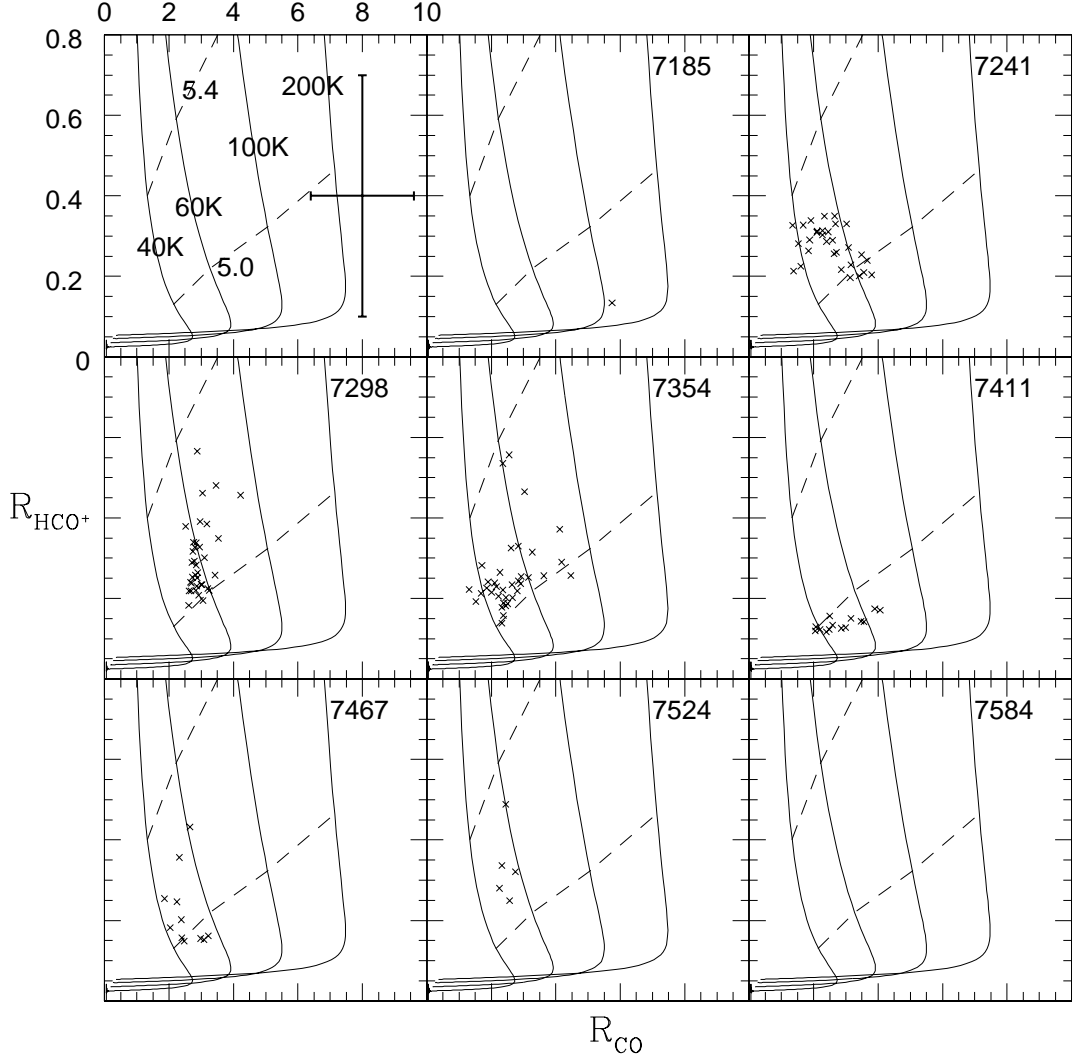}
  \caption{
    The R$_{CO}$(CO~(3--2)/CO~(1--0)) vs. R$_{HCO^+}$
    (HCO$^+$(4--3)/HCO$^+$(1--0)) plot for each 56~km~s$^{-1}$ channel.  
    The crosses show the pixel-to-pixel correlation of the two ratios.  
    The cross shown in the lower right of the first panel represents a typical
    error bar associated with the data.  
    X$_{CO}$/(dv/dr) = $10^{-7}$ (km~s$^{-1}$~pc$^{-1}$)$^{-1}$ 
    is adopted for  this particular model.  
    The solid line shows the temperature contours of 
    40, 60, 100, and 200K, and the
    dashed line shows density contours of $10^{5.0}$ and $10^{5.4}$~cm$^{-3}$.
    The lower density contours (i.e. $10^{4.0}$~cm$^{-3}$) 
    lie in the lower left corner of each panel, which are not visible 
    with the current scaling.
}
  \label{fig9}
\end{figure}
 
\begin{figure}
  \plotone{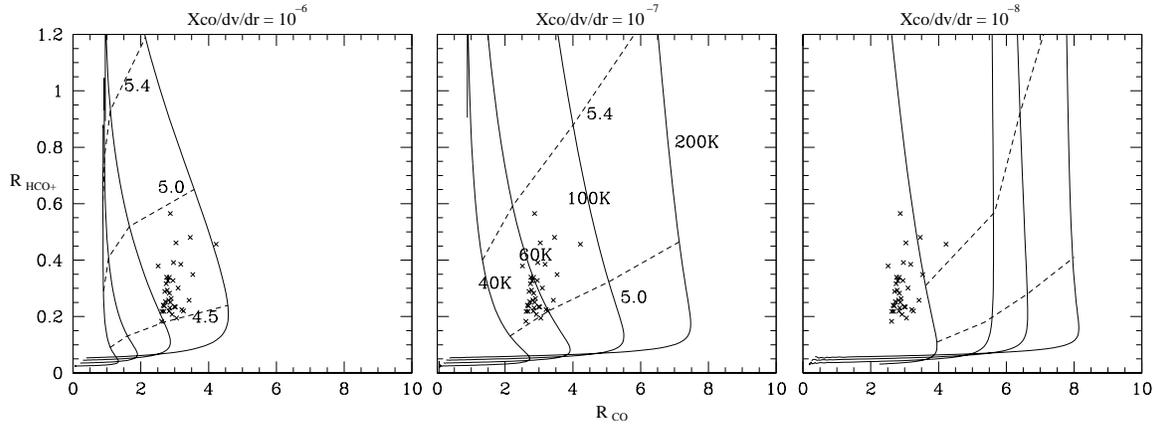}
  \caption{
    The model R$_{CO}$(CO~(3--2)/CO~(1--0)) vs. R$_{HCO^+}$
    (HCO$^+$(4--3)/HCO$^+$(1--0)) shown for different X$_{CO}$/(dv/dr) for
    comparison.  The crosses show the data points from $v=7298$~km~s$^{-1}$
    in Figure~\ref{fig9}.
    The solid line shows the temperature contours of 40, 60, 100, and 200K, 
    and the
    dashed line shows density contours of $10^{5.0}$ and $10^{5.4}$~cm$^{-3}$
    (and $10^{4.5}$~cm$^{-3}$ for 
    X$_{CO}$/(dv/dr) = $10^{-6}$~(km~s$^{-1}$~pc$^{-1}$)$^{-1}$).
    Using $X_{CO}$/(dv/dr) = $10^{-6}$~(km~s$^{-1}$~pc$^{-1}$)$^{-1}$ 
    will shift the temperature (vertical) contours leftward, 
    yielding most of the molecular gas with temperatures in excess of 200~K.  
    Using $X_{CO}$/(dv/dr) = $10^{-8}$~(km~s$^{-1}$~pc$^{-1}$)$^{-1}$
    will shift the temperature contours rightward, yielding 
    a wide range of temperatures but mostly T $< 60$~K.
}
  \label{fig10}
\end{figure}

\begin{figure}
  \plotone{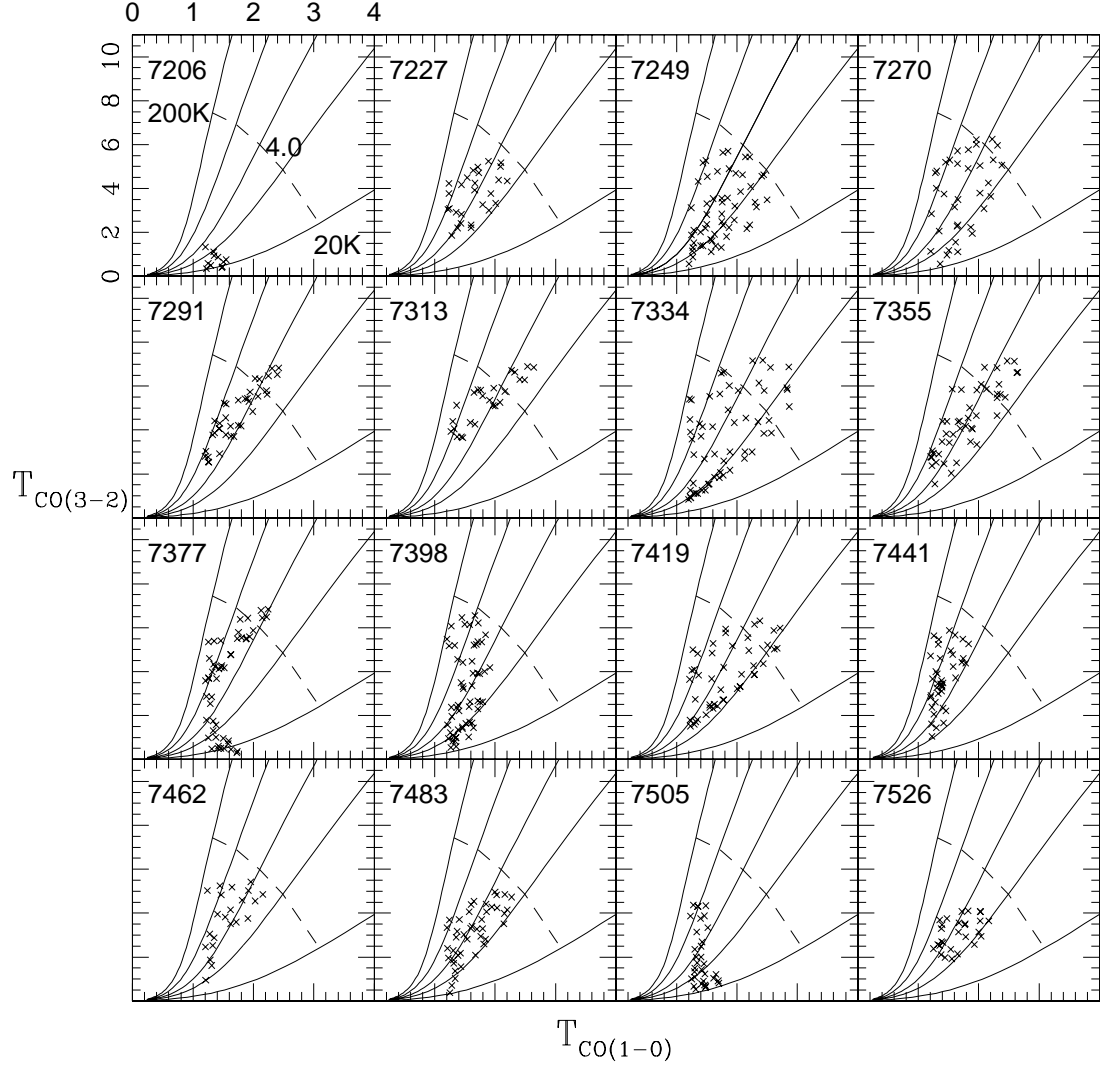}
  \caption{
    The pixel-to-pixel correlation between T$_{\rm CO(3-2)}$ 
    and T$_{\rm CO(1-0)}$ (in Kelvin) are shown in crosses, where a 
    beam filling factor of unity is assumed.  
    The solid lines show the T$_{K}=$ 20, 40, 60, 100, 200~K, and the dashed 
    line shows $n_{\rm H_2} = 10^{4.0}$~cm$^{-3}$ contour.
}
  \label{fig11}
\end{figure}

\begin{figure}
  \plotone{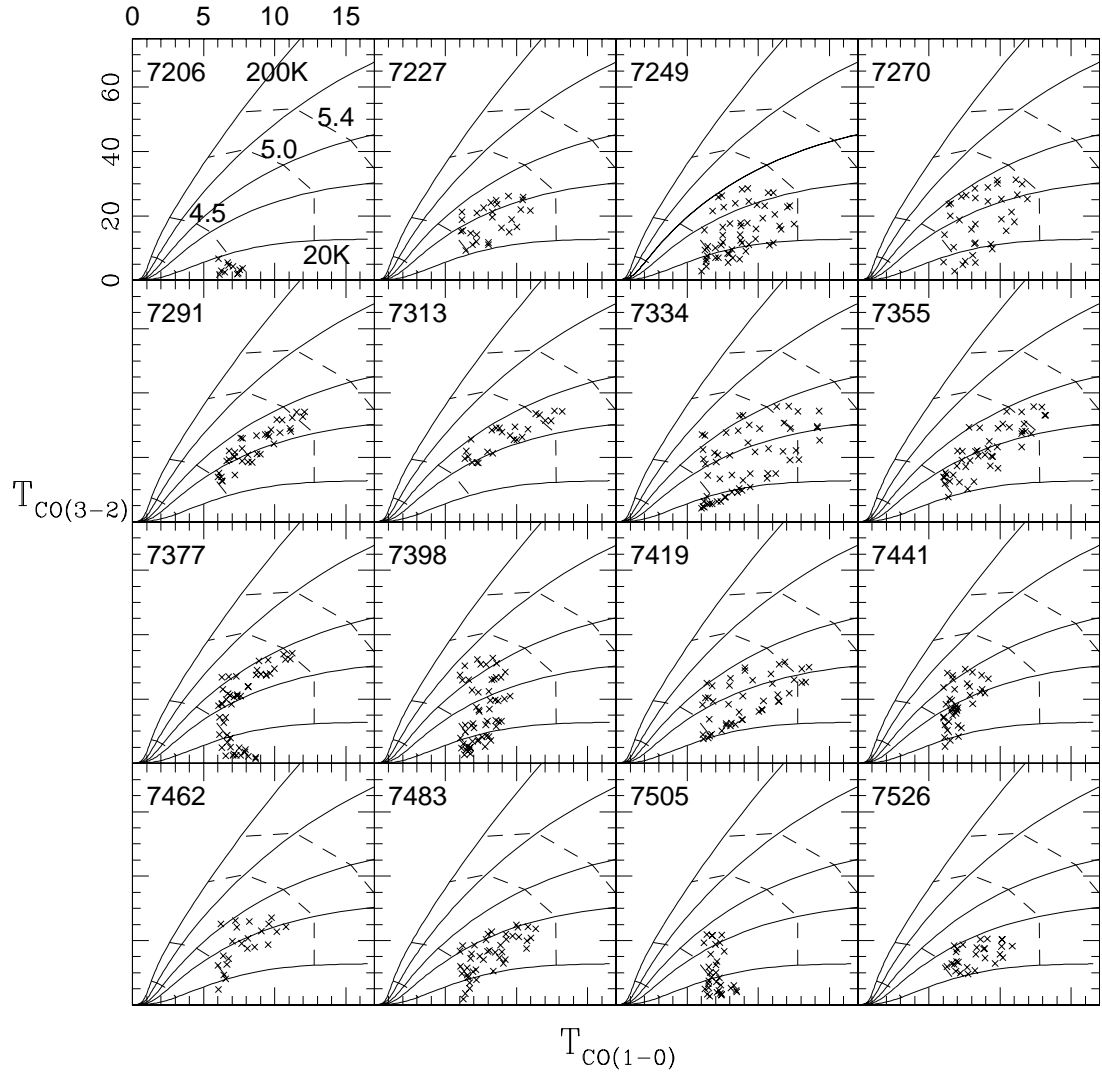}
  \caption{
    Similar to Figure~\ref{fig11} but a beam filling factor of 0.2 is 
    assumed.  
  }
  \label{fig12}
\end{figure}

\begin{figure}
  \plotone{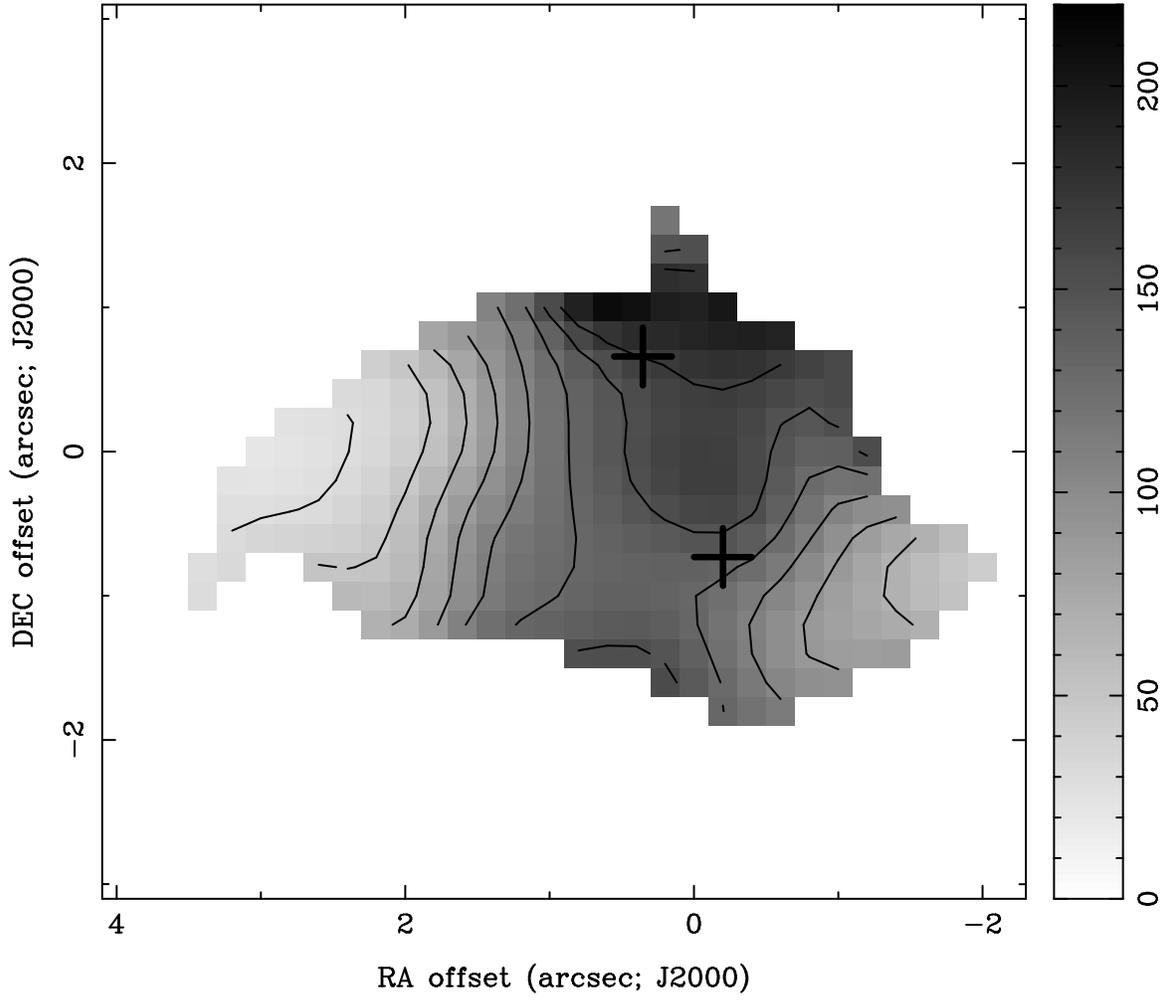}
  \caption{
    The gas-to-dust ratio map where both measurements are above 3~sigma.  
    The contours are 30,50,70,90,110,130,150,170.
    The crosses are the same as in Figure~\ref{fig4}.
}
  \label{fig13}
\end{figure}

\end{document}